\documentclass[]{jfm}

\usepackage{graphicx}
\usepackage{newtxtext}
\usepackage{newtxmath}
\usepackage{natbib}
\usepackage{hyperref}
\usepackage{caption}
\usepackage{subcaption}
\usepackage{xcolor}
\usepackage{soul}

\hypersetup{
    colorlinks = true,
    urlcolor   = blue,
    citecolor  = black,
}

\newcommand{\RomanNumeralCaps}[1]
\linenumbers

\shorttitle{Identifying invariant solutions of wall-bounded flows using variational techniques}

\shortauthor{O. Ashtari and T. M. Schneider}

\title{Identifying invariant solutions of wall-bounded three-dimensional shear flows using robust adjoint-based variational techniques}

\author{
    Omid Ashtari\aff{1}
        \and
    Tobias M. Schneider\aff{1}
        \corresp{\email{tobias.schneider@epfl.ch}}
}

\affiliation{
    \aff{1} Emergent Complexity in Physical Systems (ECPS), École Polytechnique Fédérale de Lausanne, CH-1015 Lausanne, Switzerland
}

\newcommand{\NSE}{\mathcal{F}}
\newcommand{\ug}{v}
\newcommand{\pg}{q}

\begin{document}
\maketitle
\begin{abstract}
Invariant solutions of the Navier-Stokes equations play an important role in the spatiotemporally chaotic dynamics of turbulent shear flows. Despite the significance of these solutions, their identification remains a computational challenge, rendering many solutions inaccessible and thus hindering progress towards a dynamical description of turbulence in terms of invariant solutions. 
We compute equilibria of three-dimensional wall-bounded shear flows using an adjoint-based matrix-free variational approach. 
To address the challenge of computing pressure in the presence of solid walls, we develop a formulation that circumvents the explicit construction of pressure and instead employs the influence matrix method. Together with a data-driven convergence acceleration technique based on dynamic mode decomposition, this yields a practically feasible alternative to state-of-the-art Newton methods for converging equilibrium solutions. 
We compute multiple equilibria of plane Couette flow starting from inaccurate guesses extracted from a turbulent time series. The variational method outperforms Newton(-hookstep) iterations in successfully converging from poor initial guesses, suggesting a larger convergence radius.
\end{abstract}


\section{Introduction}
Viewing fluid turbulence as a deterministic chaotic dynamical system has revealed new insights beyond what can be achieved through a purely statistical approach (see reviews by \cite{Kawahara2012} and \cite{Graham2021}).
The idea for a dynamical description by envisioning turbulence as a chaotic trajectory in the infinite-dimensional state space of the Navier-Stokes equations dates back to the seminal work of \cite{Hopf1948}.
A remarkable progress in bridging the gaps between ideas from dynamical systems theory and practically studying turbulence in this framework has been the numerical computation of \emph{invariant solutions} -- an advance that did not happen until the 1990's.
Invariant solutions are non-chaotic solutions to the governing equations with simple dependence on time. This includes equilibria (\cite{Nagata1990}), travelling waves (\cite{Faisst2003,Wedin2004}), periodic and relative periodic orbits (\cite{Kawahara2001,Chandler2013,Budanur2017a}) and invariant tori (\cite{Suri2019,Parker2023}).
In the dynamical description, the chaotic trajectory of the turbulent dynamics transiently, yet recurringly, visits the neighbourhood of the unstable invariant solutions embedded in the state space of the evolution equations. In this picture, therefore, unstable invariant solutions serve as the building blocks supporting the turbulent dynamics, and extracting them is the key for studying turbulence in the dynamical systems framework.

Equilibria of plane Couette flow (PCF) numerically computed by \cite{Nagata1990} were the first nontrivial invariant solutions discovered in a wall-bounded three-dimensional (3D) fluid flow.
Despite their lack of temporal variation, equilibrium solutions can capture essential features of chaotic flows and play an important role in characterising their chaotic dynamics.
In PCF for instance, \cite{Nagata1990}, \cite{Clever1992}, \cite{Waleffe1998}, \cite{Itano2001}, \cite{wang2007} and others compute equilibrium solutions. These equilibria typically contain wavy streaks together with pairs of staggered counter-rotating streamwise vortices, and thus capture basic structures of near-wall turbulence. \cite{Gibson2008,Gibson2009} and \cite{Halcrow2009} demonstrate how the chaotic dynamics is organised by coexisting equilibrium solutions together with their stable and unstable manifolds; \cite{Schneider2010} and \cite{Gibson2014} compute equilibria that capture localisation in the spanwise direction; \cite{Eckhardt2017} compute equilibria that capture localisation in the streamwise direction; \cite{Brand2014} compute equilibria that capture localisation in both streamwise and spanwise directions; and \cite{Reetz2019a} identify an equilibrium solution underlying self-organised oblique turbulent-laminar stripes. While equilibrium solutions have been shown to capture features of the chaotic flow dynamics, their numerical identification in very high-dimensional fluid flow problems remains challenging.

One approach to computing equilibrium solutions is to consider a \emph{root finding problem}. Irrespective of their dynamical stability, equilibria of the dynamical system $\partial u/\partial t=r(u)$ are, by definition, roots of the nonlinear operator governing the time evolution, $r(u)=0$.
The root finding problem can be solved by Newton(-Raphson) iterations.
Newton iterations are popular because of their locally quadratic convergence. However, employing Newton iterations for solving the root finding problem has two principal drawbacks:
For a system described by $N$ degrees of freedom, the update vector in each iteration is the solution to a linear system of equations whose coefficient matrix is the $N\times N$ Jacobian. Solving this large system of equations and the associated quadratically scaling memory requirement are too costly for very high-dimensional, strongly coupled fluid flow problems.
In addition to poor scaling, Newton iterations typically have a small radius of convergence, meaning that the algorithm needs to be initialised with an extremely accurate initial guess in order to converge successfully. Finding sufficiently accurate guesses is not simple even for weakly chaotic flows close to the onset of turbulence.
Newton-GMRES-hookstep is the state-of-the-art matrix-free variant of the Newton method commonly used for computing invariant solutions of fluid flows. This method defeats the $N^2$ memory scaling drawback by employing the generalised minimal residual (GMRES) method and approximating the update vector in a Krylov subspace (\cite{Saad1986,Tuckerman2019}). In addition, the robustness of the convergence is improved via hook-step trust-region optimisation (\cite{Dennis1996,Viswanath2007,Viswanath2009}).
Newton-GMRES-hookstep thereby enlarges the basin of convergence of Newton iterations. Yet, requiring an accurate initial guess is still a bottleneck of this method, and identifying unstable equilibria remains challenging.

An alternative to the root finding setup is to view the problem of computing an equilibrium solution as an \emph{optimisation problem}.
Deviation of a flow field from being an equilibrium solution can be penalised by the norm of the to-be-zeroed right-hand side operator, $\|r(u)\|$. The absolute minima of this cost function, $\|r(u)\|=0$, correspond to equilibrium solutions of the system. Therefore, the problem of finding equilibria can be recast as the minimisation of the cost function. A matrix-free method is crucial for solving this minimisation problem in very high-dimensional fluid flows. \cite{Farazmand2016} proposed an adjoint-based minimisation technique to find equilibria and travelling waves of a 2D Kolmogorov flow. The adjoint calculations allow the gradient of the cost function to be constructed analytically as an explicit function of the current flow field. This results in a matrix-free gradient-descent algorithm whose memory requirement scales linearly with the size of the problem. The adjoint-based minimisation method is significantly more robust to inaccurate initial guesses in comparison to its alternatives based on solving a root finding problem using Newton iterations. This improvement, however, is obtained by sacrificing the quadratic convergence of the Newton iterations and exhibiting slow convergence. In the context of fluid mechanics, the variational approach has been successfully applied to the 2D Kolmogorov flows (see \cite{Farazmand2016,Parker2022}). 

Despite the robust convergence and favourable scaling properties of the adjoint-based minimisation method, it has not been applied to 3D wall-bounded flows. 
Beyond the high-dimensionality of the 3D wall-bounded flows, the main challenge in the application of this method lies in handling the wall boundary conditions that cannot be imposed readily while evolving the adjoint-descent dynamics (see \textsection{\ref{sec:adjoint}}). This is in contrast to doubly periodic 2D (or triply periodic 3D) flows where the adjoint-descent dynamics is subject to periodic boundary conditions only, that can be imposed by representing variables in Fourier basis (\cite{Farazmand2016,Parker2022}).
To construct a suitable formulation for 3D flows in the presence of walls, we project the evolving velocity field onto the space of divergence-free fields and constrain pressure so that it satisfies the pressure Poisson equation instead of evolving independent of the velocity (see \textsection{\ref{sec:projection}}). 
However, solving the pressure Poisson equation with sufficient accuracy is not straightforward in wall-bounded flows.
The challenge in computing the instantaneous pressure associated with a divergence-free velocity field stems from the absence of explicit physical boundary conditions on pressure at the walls (Rempfer (2006)).
As a result, a successful implementation of the constrained dynamics hinges on resolving the challenge of accurately expressing the pressure in wall-bounded flows.

We propose an algorithm for computing equilibria of wall-bounded flows using adjoint-descent minimisation in the space of divergence-free velocity fields.
The proposed algorithm circumvents the explicit construction of pressure, thereby overcoming the inherent challenge of dealing with pressure in the application of the adjoint-descent method to wall-bounded flows.
We construct equilibria of plane Couette flow, and discuss the application of the introduced method to other wall-bounded flows and other types of invariant solutions where the challenge of dealing with pressure exists analogously.
To accelerate the convergence of the algorithm we propose a data-driven procedure which takes advantage of the almost linear behaviour of the adjoint-descent dynamics in the vicinity of an equilibrium solution. The acceleration technique approximates the linear dynamics using dynamic mode decomposition, and thereby approximates the asymptotic solution of the adjoint-descent dynamics.
The large basin of convergence together with the improved convergence properties renders the adjoint-descent method a viable alternative to the state-of-the-art Newton method.

The remainder of the manuscript is structured as follows:
The adjoint-based variational method for constructing equilibrium solutions is introduced in a general setting in \textsection{\ref{sec:adjoint_descent_general}}. 
The adjoint-descent dynamics is derived for wall-bounded shear flows in \textsection{\ref{sec:application_to_NSE}}, and an algorithm for numerically integrating the derived dynamics is presented in \textsection{\ref{sec:numerical_implementation}}.
The method is applied to plane Couette flow in \textsection{\ref{sec:results}} where the convergence of multiple equilibria is demonstrated.
The data-driven procedure for accelerating the convergence is discussed in \textsection{\ref{sec:accelerating}}.
Finally, the article is summarised and concluding remarks are provided in \textsection{\ref{sec:conclusion}}.

\section{Adjoint-descent method for constructing equilibrium solutions}
\label{sec:adjoint_descent_general}
Consider a general autonomous dynamical system
\begin{equation}
\label{eq:general_dynamical_system}
    \dfrac{\partial\mathbf{u}}{\partial t} = \mathbf{r}(\mathbf{u}),
\end{equation}
where $\mathbf{u}$ is an $n$-dimensional real-valued field defined over a $d$-dimensional spatial domain $\mathbf{x}\in\Omega\subseteq\mathbb{R}^d$ and varying with time $t\in\mathbb{R}$.
Within the space of vector fields $\mathscr{M}=\left\{\mathbf{u}:\Omega\to\mathbb{R}^n\right\}$, the evolution of $\mathbf{u}$ is governed by the smooth nonlinear operator $\mathbf{r}$ subject to time-independent boundary conditions at $\partial\Omega$, the boundary of $\Omega$.
Equilibrium solutions of this dynamical system are $\mathbf{u}^\ast\in\mathscr{M}$ for which
\begin{equation}
\label{eq:def_equilibrium_general}
    \mathbf{r}(\mathbf{u}^\ast) = \mathbf{0}.
\end{equation}
The residual of Equation \eqref{eq:def_equilibrium_general} is not zero for non-equilibrium states $\mathbf{u}\neq\mathbf{u}^\ast$. We thus penalise non-equilibrium states by the non-negative cost function $J^2$ defined as
\begin{equation}
    J^2 = \left<\mathbf{r}(\mathbf{u}),\mathbf{r}(\mathbf{u})\right>,
\end{equation}
where $\left<\cdot,\cdot\right>$ denotes an inner product defined on $\mathscr{M}$. The cost function takes zero value if and only if $\mathbf{u}=\mathbf{u}^\ast$. We thereby recast the problem of finding equilibrium solutions $\mathbf{u}^\ast$ as a minimisation problem over $\mathscr{M}$, and look for the global minima of $J^2$ at which $J^2=0$, following the arguments of \cite{Farazmand2016}.

In order to find minima of $J^2$, we construct another dynamical system in $\mathscr{M}$ whose evolution monotonically decreases the cost function $J^2$. The objective is to define an evolution equation
\begin{equation}
\label{eq:adjoint_descent_general}
    \dfrac{\partial\mathbf{u}}{\partial\tau}=\mathbf{g}(\mathbf{u}),
\end{equation}
where the choice of the operator $\mathbf{g}$ guarantees
\begin{equation}
    \dfrac{\partial J^2}{\partial\tau}\leq 0\,; \quad \forall\tau.
\end{equation}
Here, $\tau$ is a fictitious time that parametrizes the evolution governed by the constructed dynamics. The rate of change of $J^2$ along trajectories of the dynamical system \eqref{eq:adjoint_descent_general} is
\begin{equation}
\label{eq:rate_of_J_general}
    \dfrac{\partial J^2}{\partial\tau} = 2\left<\mathscr{L}(\mathbf{u};\mathbf{g}),\mathbf{r}(\mathbf{u})\right>,
\end{equation}
where $\mathscr{L}(\mathbf{u};\mathbf{g})$ is the directional derivative of $\mathbf{r}(\mathbf{u})$ along $\partial\mathbf{u}/\partial\tau=\mathbf{g}$:
\begin{equation}
    \mathscr{L}(\mathbf{u};\mathbf{g}) = \lim_{\epsilon\to0}\dfrac{\mathbf{r}(\mathbf{u}+\epsilon\mathbf{g})-\mathbf{r}(\mathbf{u})}{\epsilon}.
\end{equation}
We can rewrite Equation \eqref{eq:rate_of_J_general} as
\begin{equation}
    \dfrac{\partial J^2}{\partial\tau} = 2\left<\mathscr{L}^\dagger(\mathbf{u};\mathbf{r}),\mathbf{g}(\mathbf{u})\right>,
\end{equation}
where $\mathscr{L}^\dagger$ is the adjoint operator of the directional derivative $\mathscr{L}$, with the following definition:
\begin{equation}
    \big<\mathscr{L}(\mathbf{v};\mathbf{v}'),\mathbf{v}''\big>=\big<\mathscr{L}^\dagger(\mathbf{v};\mathbf{v}''),\mathbf{v}'\big> \,; \quad \forall\; \mathbf{v},\mathbf{v}',\mathbf{v}''\in\mathscr{M}.
\end{equation}
To guarantee the monotonic decrease of $J^2$ with $\tau$ we choose
\begin{equation}
    \label{eq:adj_rhs_general}
    \mathbf{g}(\mathbf{u}) = -\mathscr{L}^\dagger(\mathbf{u};\mathbf{r}).
\end{equation}
This choice results in monotonic decrease of $J^2$ along solution trajectories of the adjoint dynamical system \eqref{eq:adjoint_descent_general}:
\begin{equation}
    \dfrac{\partial J^2}{\partial\tau} = -2\left<\mathscr{L}^\dagger(\mathbf{u};\mathbf{r}),\mathscr{L}^\dagger(\mathbf{u};\mathbf{r})\right>\leq 0.
\end{equation}

\begin{figure}
     \centering
     \begin{subfigure}[b]{0.38\textwidth}
         \centering
         \includegraphics[height = 4.75cm]{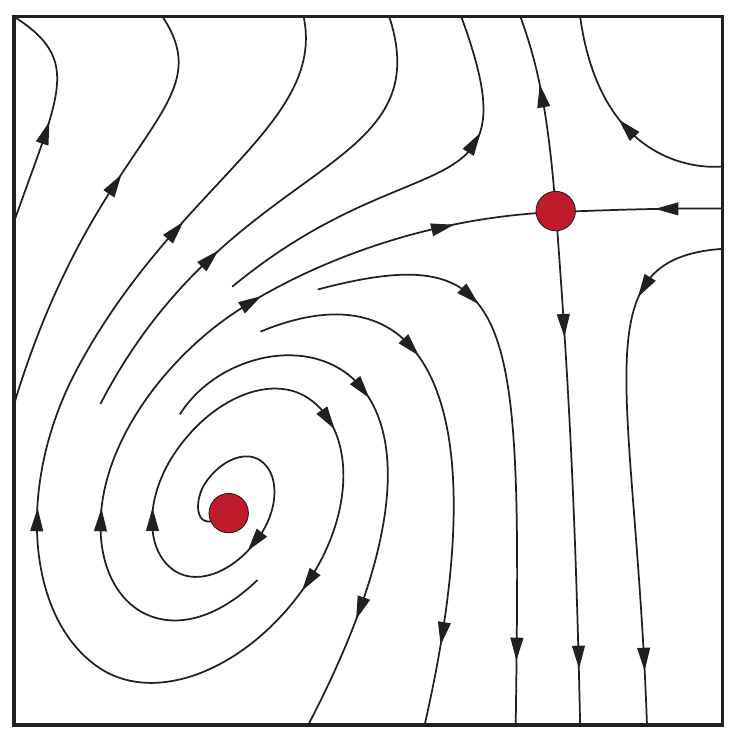}
         \caption{$\partial\mathbf{u}/\partial t=\mathbf{r}(\mathbf{u})$}
         \label{fig:idea_sub_original}
     \end{subfigure}
     \begin{subfigure}[b]{0.38\textwidth}
         \centering
         \includegraphics[height = 4.77cm]{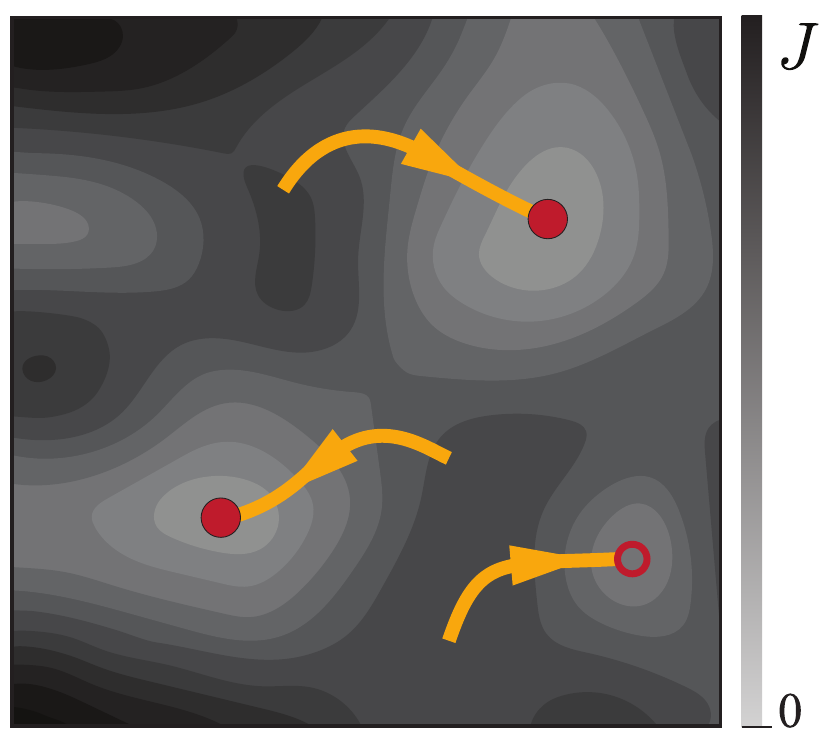}
         \caption{$\partial\mathbf{u}/\partial \tau=\mathbf{g}(\mathbf{u})$}
         \label{fig:idea_sub_adjoint}
     \end{subfigure}
     \caption{Replacing the original dynamics with the gradient descent of the cost function $J=\|\mathbf{r}(\mathbf{u})\|$ by the adjoint-descent method. Panel (a) schematically shows the trajectories and two equilibria of the original system parametrized by the physical time $t$, while panel (b) shows contours of $J$ and sample trajectories of its gradient flow parametrized by the fictitious time $\tau$. Trajectories of the adjoint-descent dynamics converge to a stable fixed point, that is either an equilibrium of the original dynamics, where the global minimum value of $J=0$ is achieved, or a state at which $J$ takes a local minimum value.}
     \label{fig:idea}
\end{figure}

In summary, in order to find equilibria of $\partial\mathbf{u}/\partial t=\mathbf{r}(\mathbf{u})$ the variational approach proposed by \cite{Farazmand2016} constructs a globally contracting dynamical system $\partial\mathbf{u}/\partial\tau=\mathbf{g}(\mathbf{u})$, that is essentially the gradient descent of the cost function $J^2$.
Every trajectory of the constructed dynamical system eventually reaches a stable equilibrium corresponding to a minimum of the cost function. Equilibria of the original dynamics are equilibria of the adjoint dynamics at which the cost function takes its global minimum value $J^2=0$. However, the adjoint dynamics might have other equilibria that correspond to a local minimum of the cost function with $J^2>0$, and are not equilibria of the original dynamics. This is schematically illustrated in Figure \ref{fig:idea}.
Finding equilibria of $\partial\mathbf{u}/\partial t=\mathbf{r}(\mathbf{u})$ requires integrating the adjoint dynamics $\partial\mathbf{u}/\partial\tau=\mathbf{g}(\mathbf{u})$ forward in the fictitious time $\tau$. The solutions obtained at $\tau\to\infty$ for which $J^2=0$ are equilibria of the original system. Otherwise, when the trajectory gets stuck in a local minimum of the cost function, the search fails and the adjoint dynamics should be integrated from another initial condition.

\section{Application to the wall-bounded shear flows}
\label{sec:application_to_NSE}
\subsection{Governing equations}
\label{sec:governing_equations}
We consider the flow in a three-dimensional rectangular domain $\Omega$ of non-dimensional size $x\in[0,L_x)$, $y\in[-1,+1]$ and $z\in[0,L_z)$. The domain is bounded in $y$ between two parallel plates, and is periodic in the lateral directions $x$ and $z$. Incompressible, isotherm flow of a Newtonian fluid is governed by the Navier-Stokes equations (NSE). The non-dimensional, perturbative form of the NSE reads
\begin{gather}
\label{eq:momentum}
    \frac{\partial\mathbf{u}}{\partial t}=-\left[(\mathbf{u}_b\cdot\nabla)\mathbf{u}+(\mathbf{u}\cdot\nabla)\mathbf{u}_b+(\mathbf{u}\cdot\nabla)\mathbf{u}\right]-\nabla p+\frac{1}{Re}\Delta\mathbf{u}=:\NSE(\mathbf{u},p),\\
\label{eq:continuity}
    \nabla\cdot\mathbf{u}=0.
\end{gather}
Here, $Re$ is the Reynolds number and $\mathbf{u}_b$ is the laminar base flow velocity field. The fields $\mathbf{u}$ and $p$ are the deviations of the total velocity and pressure from the base flow velocity and pressure fields, respectively.
For common driving mechanisms such as the motion of walls in the $xz$ plane, externally imposed pressure differences, or injection/suction through the walls, the laminar base flow satisfies the inhomogeneous boundary conditions (BCs), absorbs body forces, and is known a priori.
Consequently, the perturbative Navier-Stokes equations \eqref{eq:momentum} and \eqref{eq:continuity} are subject to the boundary conditions
\begin{gather}
    \label{eq:NSE_BC1}
    \mathbf{u}(x,y=\pm1,z;t)=\mathbf{0},\\
    \label{eq:NSE_BC2}
    [\mathbf{u},p](x=0,y,z;t)=[\mathbf{u},p](x=L_x,y,z;t),\\
    \label{eq:NSE_BC3}
    [\mathbf{u},p](x,y,z=0;t)=[\mathbf{u},p](x,y,z=L_z;t).
\end{gather}
The canonical wall-bounded shear flows such as plane Couette flow, plane Poiseuille flow and asymptotic suction boundary layer flow are governed by the incompressible NSE \eqref{eq:momentum}-\eqref{eq:NSE_BC3} where $\mathbf{u}_b$ differentiates them from one another. We derive the adjoint-descent dynamics based on a general base flow velocity field $\mathbf{u}_b$, and in \textsection{\ref{sec:results}} demonstrate the adjoint-based method for the specific case of plane Couette flow.

The state space $\mathscr{M}$ of the NSE contains velocity fields $\mathbf{u}:\Omega\to\mathbb{R}^3$ of zero divergence which satisfy the kinematic conditions \eqref{eq:NSE_BC1}-\eqref{eq:NSE_BC3}. The space $\mathscr{M}$ carries the standard energy-based $L_2$ inner product denoted with $\left<\cdot,\cdot\right>_\mathscr{M}$.
The pressure $p$ associated with an admissible velocity field $\mathbf{u}\in\mathscr{M}$ ensures that under the NSE dynamics the velocity remains divergence-free,
\begin{equation}
\label{eq:physical_pressure_eqn}
    \partial(\nabla\cdot\mathbf{u})/\partial t = \nabla\cdot\NSE(\mathbf{u},p) = 0,
\end{equation}
while remaining compatible with the no-slip boundary conditions \eqref{eq:NSE_BC1},
\begin{equation}
\label{eq:physical_pressure_BCs}
    \partial\mathbf{u}/\partial t\big|_{y=\pm1} = \NSE(\mathbf{u},p)\big|_{y=\pm1} = \mathbf{0}.
\end{equation}
This requires $p$ to satisfy the Poisson equation with a velocity-dependent source term (\cite{Canuto2007,Rempfer2006}).

We could not derive a variational dynamics based on expressing pressure explicitly as the solution to this Poisson equation.
Therefore, instead of the state space $\mathscr{M}$ of the NSE we define the search space such that `velocity' and `pressure' can evolve independently. Accordingly, we define the cost function such that residuals of both Equations \eqref{eq:momentum} and \eqref{eq:continuity} are included. Otherwise, the derivation follows \textsection{\ref{sec:adjoint_descent_general}}.

\subsection{The search space}
We define the inner product space of general flow fields as
\begin{equation}
    \mathscr{P}=\left\{\begin{bmatrix}
        \mathbf{\ug}\\
        \pg
    \end{bmatrix}\left|
    \begin{array}{l}
        \mathbf{\ug}:\Omega \rightarrow \mathbb{R}^3\\
        \pg:\Omega \rightarrow \mathbb{R}\\
        \mathbf{\ug} \text{ and } \pg \text{ periodic in } x \text{ and } z
    \end{array}\right.\right\},
    \label{eq:def_general_flow_fields}
\end{equation}
where $\mathbf{\ug}$ and $\pg$ are sufficiently smooth functions of space. Hereafter, the symbols $\mathbf{u},p$ indicate physically admissible velocity and pressure, implying $\mathbf{u}\in\mathscr{M}$ and $p$ satisfying the relevant Poisson equation.
The space of general flow fields $\mathscr{P}$ is endowed with the real-valued inner product
\begin{equation}
\label{eq:inner_product}
    \begin{gathered}
        \left<\ ,\ \right>:\mathscr{P} \times \mathscr{P}\rightarrow \mathbb{R},\\
        \left<
        \begin{bmatrix}
            \mathbf{\ug}_1\\
            \pg_1
        \end{bmatrix}
        ,
        \begin{bmatrix}
            \mathbf{\ug}_2\\
            \pg_2
        \end{bmatrix}
        \right>=\int_\Omega{\left(\mathbf{\ug}_1\cdot\mathbf{\ug}_2+\pg_1\pg_2\right)}\text{d}\mathbf{x}.
    \end{gathered}
\end{equation}
Here $\cdot$ is the conventional Euclidean inner product in $\mathbb{R}^3$.
Physically admissible velocity and pressure fields form the following subset of the general flow fields:
\begin{equation}
    \mathscr{P}_p=\left\{
        \begin{bmatrix}
            \mathbf{u}\\
            p
        \end{bmatrix}\in\mathscr{P}_0\left|
        \begin{array}{l}
        \nabla\cdot\mathbf{u}=0\\
        \nabla\cdot\NSE(\mathbf{u},p) = 0\\
        \NSE(\mathbf{u},p)\big|_{y=\pm1} = \mathbf{0}
    \end{array}\right.\right\},
\end{equation}
where $\mathscr{P}_0$ is the subset of $\mathscr{P}$ whose vector-valued component satisfies the homogeneous Dirichlet BC at the walls:
\begin{equation}
        \mathscr{P}_0=\left\{
        \begin{bmatrix}
            \mathbf{\ug}\\
            \pg
        \end{bmatrix}\in\mathscr{P}\;\bigg|\;
        \mathbf{\ug}(y=\pm1)=\mathbf{0}\right\}.
        \label{eq:def_Dirichlet_sub_general_flow_fields}
\end{equation}

Equilibrium solutions of the NSE are $[\mathbf{u}^\ast,p^\ast]\in\mathscr{P}_p$ for which
\begin{equation}
\label{eq:def_fixed_point}
    \NSE(\mathbf{u}^\ast,p^\ast)=\mathbf{0}.
\end{equation}
We aim to impose the zero-divergence constraint together with the defining property of an equilibrium solution via the variational minimisation discussed in \textsection{\ref{sec:adjoint_descent_general}}.
To that end, we consider an evolution in the space of general flow fields $\mathbf{U}=[\mathbf{\ug},\pg]\in\mathscr{P}_0$ in which the velocity and the pressure component are evolved independently.
A flow field $\mathbf{U}\in\mathscr{P}_0$ neither necessarily satisfies the defining property of an equilibrium solution nor the zero-divergence constraint.
Therefore, we define the residual field $\mathbf{R}\in\mathscr{P}$ associated with a general flow field as
\begin{equation}
\label{eq:residual_def}
    \mathbf{R}=\begin{bmatrix}
        \mathbf{r}_1\\
        r_2
    \end{bmatrix}=\begin{bmatrix}
        \NSE(\mathbf{\ug},\pg)\\
        \nabla\cdot\mathbf{\ug}
    \end{bmatrix},
\end{equation}
and the cost function $J^2$ as
\begin{equation}
\label{eq:cost_NSE}
    J^2 = \int_\Omega\left(\NSE^2(\mathbf{\ug},\pg)+\left(\nabla\cdot\mathbf{\ug}\right)^2\right)\text{d}\mathbf{x} = \int_\Omega\left(\mathbf{r}_1\cdot\mathbf{r}_1+r_2^2\right)\text{d}\mathbf{x} = \left<\mathbf{R},\mathbf{R}\right>.
\end{equation}
At the global minima of the cost function, $J^2=0$, the defining property of an equilibrium solution \eqref{eq:def_fixed_point} and the incompressibility constraint \eqref{eq:continuity} are both satisfied. 
The operator $\mathbf{G}=[\mathbf{g}_1,g_2]$ acting on general flow fields $\mathbf{U}=[\mathbf{\ug},\pg]\in\mathscr{P}_0$ is constructed such that an equilibrium solution $[\mathbf{u}^\ast,p^\ast]$ is obtained by evolving the variational dynamics
\begin{equation}
\label{eq:variational_dynamics_general_flow_fields}
    \dfrac{\partial\mathbf{U}}{\partial\tau} = \dfrac{\partial}{\partial\tau}
    \begin{bmatrix}
        \mathbf{\ug}\\
        \pg
    \end{bmatrix}=
    \begin{bmatrix}
        \mathbf{g}_1\\
        g_2
    \end{bmatrix}.
\end{equation}
The operator $\mathbf{G}$ is derived following the adjoint-based method described in \textsection{\ref{sec:adjoint_descent_general}} to guarantee the monotonic decrease of the cost function along trajectories of the variational dynamics \eqref{eq:variational_dynamics_general_flow_fields}.

\subsection{Adjoint operator for the NSE}
\label{sec:adjoint}
The variational dynamics \eqref{eq:variational_dynamics_general_flow_fields} must ensure that the flow field $\mathbf{U}$ remains within $\mathscr{P}_0$, thus $\mathbf{U}$ is periodic in $x$ and $z$, and its velocity component $\mathbf{\ug}$ takes zero value at the walls for all $\tau$.
In order for these properties of $\mathbf{U}$ to be preserved under the variational dynamics, the operator $\mathbf{G}$ must be periodic in $x$ and $z$, and $\mathbf{g}_1=\partial\mathbf{\ug}/\partial\tau$ must take zero value at the walls, meaning that $\mathbf{G}\in\mathscr{P}_0$.
In addition, we choose the residual $\mathbf{R}$ to lie within $\mathscr{P}_0$. The periodicity of $\mathbf{R}$ in $x$ and $z$ automatically results from the spatial periodicity of $\mathbf{U}$ in these two directions. However, at the walls we enforce the condition $\mathbf{r}_1(\mathbf{\ug},\pg)|_{y=\pm 1}=\NSE(\mathbf{\ug},\pg)|_{y=\pm 1}=\mathbf{0}$.
With the choice of $\mathbf{U},\mathbf{R},\mathbf{G}\in\mathscr{P}_0$, the flow field remains within $\mathscr{P}_0$ as desired. Following this choice, all the boundary terms resulting from partial integrations in the derivation of the adjoint operator cancel out (see Appendix \ref{app:adjoint_derivation}), and the adjoint of the directional derivative of $\mathbf{R}(\mathbf{U})$ along $\mathbf{G}$ is obtained as
\begin{align}
    \mathscr{L}^\dagger_1 &= \left(\nabla\mathbf{r}_1\right)\left(\mathbf{u}_b+\mathbf{\ug}\right)-\left(\nabla(\mathbf{u}_b+\mathbf{\ug})\right)^\top\mathbf{r}_1+\dfrac{1}{Re}\Delta\mathbf{r}_1+r_2\mathbf{r}_1-\nabla r_2, \\ 
    \mathscr{L}^\dagger_2 & = \nabla\cdot\mathbf{r}_1.
\end{align}
Therefore, with $\mathbf{G}=-\mathscr{L}^\dagger(\mathbf{U};\mathbf{R})$ the variational dynamics takes the form
\begin{align}
    \label{eq:rate_of_u}
    \dfrac{\partial\mathbf{\ug}}{\partial\tau}=&-\mathscr{L}^\dagger_1 = -\left(\nabla\mathbf{r}_1\right)\left(\mathbf{u}_b+\mathbf{\ug}\right)+\left(\nabla(\mathbf{u}_b+\mathbf{\ug})\right)^\top\mathbf{r}_1-\dfrac{1}{Re}\Delta\mathbf{r}_1-r_2\mathbf{r}_1+\nabla r_2, \\ 
    \label{eq:rate_of_P}
    \dfrac{\partial \pg}{\partial \tau}= &-\mathscr{L}^\dagger_2 = -\nabla\cdot\mathbf{r}_1.
\end{align}
Equation \eqref{eq:rate_of_u} is fourth order with respect to $\mathbf{\ug}$ and Equation \eqref{eq:rate_of_P} is second order with respect to $\pg$. Therefore, four BCs for each component of $\mathbf{\ug}$ and two BCs for $\pg$ are required in the inhomogeneous direction $y$.
The choice of $\mathbf{U}\in\mathscr{P}_0$ implies $\mathbf{\ug}=\mathbf{0}$ and the choice of $\mathbf{R}\in\mathscr{P}_0$ implies $\mathbf{r}_1=\NSE(\mathbf{\ug},\pg)=\mathbf{0}$ at each wall. Consequently, the adjoint-descent dynamics requires two additional wall BCs in order to be well-posed.
As the additional BCs, we impose $\nabla\cdot\mathbf{\ug}=0$ at each wall.
Therefore, the adjoint-descent dynamics is subject to the following BCs:
\begin{gather}
    \label{eq:adj_periodicx_BC}
    [\mathbf{\ug},\pg](x=0,y,z;\tau)=[\mathbf{\ug},\pg](x=L_x,y,z;\tau),\\
    \label{eq:adj_periodicz_BC}
    [\mathbf{\ug},\pg](x,y,z=0;\tau)=[\mathbf{\ug},\pg](x,y,z=L_z;\tau),\\
    \label{eq:adj_noslip_BC}
    \mathbf{\ug}(x,y=\pm1,z;\tau)=\mathbf{0},\\
    \label{eq:adj_P_BC}
    \left[-\mathbf{u}_b\cdot\nabla\mathbf{\ug} - \nabla \pg+\frac{1}{Re}\Delta\mathbf{\ug} \right]_{y=\pm1}=\mathbf{0},\\
    \label{eq:adj_div_BC} \left.\mathbf{e}_y\cdot\dfrac{\partial\mathbf{\ug}}{\partial y}\right|_{y=\pm1} = 0,
\end{gather}
where the BC \eqref{eq:adj_P_BC} is $\mathbf{r}_1=\NSE(\mathbf{\ug},\pg)=\mathbf{0}$ and the BC \eqref{eq:adj_div_BC} is $\nabla\cdot\mathbf{\ug}=0$ at the walls obtained by substituting $\mathbf{\ug}(y=\pm1)=\mathbf{0}$ in the definition of $\NSE(\mathbf{\ug},\pg)$ and $\nabla\cdot\mathbf{\ug}$, respectively.
The choice of the additional BCs is consistent with the properties of the state space of the NSE and physically meaningful. Note, however, that the BC \eqref{eq:adj_div_BC} does not need to be explicitly enforced during the derivation of the adjoint operator.
In the absence of solid walls in a doubly periodic 2D or a triply periodic 3D domain, the BCs \eqref{eq:adj_noslip_BC}-\eqref{eq:adj_div_BC} do not apply. Instead, the fields are subject to periodic BCs only.

Numerically imposing the BCs \eqref{eq:adj_noslip_BC}-\eqref{eq:adj_div_BC} while evolving Equations \eqref{eq:rate_of_u} and \eqref{eq:rate_of_P} forward in the fictitious time is not straightforward.
Consequently, instead of advancing the derived variational dynamics directly, we constrain the adjoint-descent dynamics to the subset of physical flow fields $\mathscr{P}_p$. 
Within this subset, pressure does not evolve independently but satisfies the pressure Poisson equation. Thereby, we obtain an evolution equation for the velocity within the state space of the NSE.
This allows us to employ the influence matrix method (\cite{Kleiser1980}) to integrate the constrained adjoint-descent dynamics.

\subsection{Variational dynamics constrained to the subset of physical flow fields}
\label{sec:projection}
To obtain a numerically tractable variational dynamics, we constrain the adjoint-descent dynamics \eqref{eq:rate_of_u}-\eqref{eq:adj_div_BC} to the subset of physical flow fields $\mathscr{P}_p$.
Within $\mathscr{P}_p$, the velocity component $\mathbf{u}$ is divergence-free over the entire domain. In addition, the pressure component $p$ is no longer governed by an explicit evolution equation, but by Poisson equation with a velocity-dependent source term.
Let $p = \mathcal{P}[\mathbf{u}]$ denote the solution to the Poisson equation yielding pressure associated with an instantaneous divergence-free velocity $\mathbf{u}$.
In order for $\mathbf{u}$ to remain divergence-free, $\mathbf{g}_1$ needs to be projected onto the space of divergence-free fields, yielding the evolution
\begin{equation}
\label{eq:projection}
    \dfrac{\partial\mathbf{u}}{\partial\tau} = \mathbb{P}\left\{ -\left(\nabla\mathbf{r}_1\right)\left(\mathbf{u}_b+\mathbf{u}\right)+\left(\nabla(\mathbf{u}_b+\mathbf{u})\right)^\top\mathbf{r}_1-\dfrac{1}{Re}\Delta\mathbf{r}_1\right\}=:\mathbf{f},
\end{equation}
where $\mathbb{P}$ denotes the projection operator. The argument of the operator $\mathbb{P}$ is the right-hand side of Equation \eqref{eq:rate_of_u} with $r_2=0$ and $\nabla r_2=\mathbf{0}$ that result from the zero divergence of $\mathbf{u}$.
According to the Helmholtz's theorem, a smooth 3D vector field can be decomposed into a divergence-free and a curl-free component. Thus, $\mathbf{g}_1=\partial\mathbf{u}/\partial\tau$ is decomposed as $\mathbf{g}_1=\mathbf{f} - \nabla \phi$ where $\mathbf{f}=\mathbb{P}\left\{\mathbf{g}_1\right\}$ is the divergence-free component and $\phi$ is the scalar potential whose gradient gives the curl-free component. Therefore, the evolution of the divergence-free velocity is governed by
\begin{gather}
\label{eq:adjoint_evolution_divergence_free}
    \dfrac{\partial\mathbf{u}}{\partial\tau} = -\left(\nabla\mathbf{r}_1\right)\left(\mathbf{u}_b+\mathbf{u}\right)+\left(\nabla(\mathbf{u}_b+\mathbf{u})\right)^\top\mathbf{r}_1+\nabla\phi-\dfrac{1}{Re}\Delta\mathbf{r}_1,\\
    \nabla\cdot\mathbf{u}=0,\label{eq:adj_zero_divergence}
\end{gather}
subject to
\begin{gather}
    \mathbf{u}(x=0,y,z;\tau)=\mathbf{u}(x=L_x,y,z;\tau),\label{eq:adj_periodicx_BC_}\\
    \mathbf{u}(x,y,z=0;\tau)=\mathbf{u}(x,y,z=L_z;\tau),\label{eq:adj_periodicz_BC_}\\
    \mathbf{u}(x,y=\pm1,z;\tau)=\mathbf{0},\label{eq:adj_noslip_BC_}\\
    \mathbf{r}_1(x,y=\pm1,z;\tau)=\mathbf{0},\label{eq:adj_P_BC_}
\end{gather}
where $\mathbf{r}_1=\mathbf{r}_1(\mathbf{u},\mathcal{P}[\mathbf{u}])$ and thus the BC \eqref{eq:adj_P_BC_} is automatically satisfied.
It is necessary to verify that the constrained variational dynamics still guarantees a monotonic decrease of the cost function.
For $\mathbf{U}\in\mathscr{P}_p$, the scalar component of the steepest descent direction, $g_2$, vanishes (see Equation \eqref{eq:rate_of_P}). Therefore, according to the definition of the inner product on $\mathscr{P}_p$ \eqref{eq:inner_product}, it is sufficient to verify that $\int_\Omega(\mathbf{f}\cdot\mathbf{g}_1)\,\mathrm{d}\mathbf{x}=\left<\mathbf{f},\mathbf{g}_1\right>_\mathscr{M}\geq 0$. The Helmholtz decomposition is an orthogonal decomposition with respect to the $L_2$ inner product defined on the state space of the NSE, $\left<\mathbf{f},\nabla\phi\right>_\mathscr{M}=
0$.
Therefore, $\left<\mathbf{f},\mathbf{g}_1\right>_\mathscr{M}=\left<\mathbf{f},\mathbf{f}\right>_\mathscr{M}-\left<\mathbf{f},\nabla\phi\right>_\mathscr{M}=\|\mathbf{f}\|^2_\mathscr{M}\geq 0$, and thus the evolution of $\mathbf{u}$ along $\mathbf{f}$ guarantees the monotonic decrease of the cost function, as desired.

The variational dynamics \eqref{eq:adjoint_evolution_divergence_free}-\eqref{eq:adj_P_BC_} is equivariant under continuous translations in the periodic directions $x$ and $z$.
Furthermore, one can verify through simple calculations that this dynamics is also equivariant under the action of any reflection or rotation permitted by the laminar base velocity field $\mathbf{u}_b$.
Consequently, the symmetry group generated by translations, reflections and rotations in the obtained variational dynamics is identical to that of the NSE \eqref{eq:momentum}-\eqref{eq:NSE_BC3}. Therefore, to construct equilibria within a particular symmetry-invariant subspace of the NSE, one can use initial conditions from the same symmetry-invariant subspace to initialise the variational dynamics, and the variational dynamics preserves the symmetries of the initial condition.

In the variational dynamics, the scalar field $\phi$ plays a role analogous to the pressure $p$ in the incompressible NSE. The scalar fields $\phi$ and $p$ adjust themselves to the instantaneous physical velocity $\mathbf{u}$ such that $\nabla\cdot\mathbf{u}=0$ and $\mathbf{u}(y=\pm1)=\mathbf{0}$ are preserved under the evolution with the fictitious time $\tau$ and the physical time $t$, respectively.
Similar to the pressure in the NSE, $\phi$ satisfies a Poisson equation with a velocity-dependent source term.
Solving the Poisson equation for $\phi$ and $p$ is a numerically challenging task in the present wall-bounded configuration (\cite{Rempfer2006}).
Therefore, instead of attempting to compute $p$ and $\phi$ and thereby advancing the variational dynamic \eqref{eq:adjoint_evolution_divergence_free}, we formulate the numerical integration scheme based on the influence matrix method (\cite{Kleiser1980}) where the no-slip BC and zero divergence are precisely satisfied while the explicit construction of $p$ and $\phi$ is circumvented.

\section{Numerical implementation}
\label{sec:numerical_implementation}
To advance the variational dynamics \eqref{eq:adjoint_evolution_divergence_free}-\eqref{eq:adj_P_BC_} without explicitly computing $\phi$ and $p$, we take advantage of the structural similarity between the variational dynamics and the NSE.
In order to evaluate the right-hand side of Equation \eqref{eq:adjoint_evolution_divergence_free}, we consider the following PDE for the residual field $\mathbf{r}_1$:
\begin{equation}
\label{eq:g_trick}
    \dfrac{\partial\mathbf{r}_1}{\partial\hat{\tau}}=-\left(\mathbf{N}(\mathbf{r}_1)-\nabla\phi+\dfrac{1}{Re}\Delta\mathbf{r}_1\right),
\end{equation}
subject to 
\begin{gather}
\label{eq:g_trick_no_slip}
    \mathbf{r}_1(y=\pm1)=\mathbf{0},\\
\label{eq:g_trick_zero_divergence}
    \nabla\cdot\mathbf{r}_1=0,
\end{gather}
where $\mathbf{N}(\mathbf{r}_1)=\left(\nabla\mathbf{r}_1\right)\left(\mathbf{u}_b+\mathbf{u}\right)-\left(\nabla(\mathbf{u}_b+\mathbf{u})\right)^\top\mathbf{r}_1$ with both $\mathbf{u}$ and $\mathbf{u}_b$ being treated as constant fields. We use the dummy Equation \eqref{eq:g_trick} to evaluate the right-hand side of Equation \eqref{eq:adjoint_evolution_divergence_free} since the instantaneously evaluated right-hand side of these two systems are identically equal.
For brevity, we are omitting the periodic BCs in $x$ and $z$ since spatial periodicity can be enforced via spectral representation in an appropriate basis, such as Fourier basis, that is periodic by construction.
Equation \eqref{eq:g_trick} together with the BC \eqref{eq:g_trick_no_slip} and the zero-divergence constraint \eqref{eq:g_trick_zero_divergence} resembles the structure of the incompressible NSE:
\begin{equation}
\label{eq:u_trick}
    \dfrac{\partial\mathbf{u}}{\partial t}=\mathbf{M}(\mathbf{u})-\nabla p+\dfrac{1}{Re}\Delta\mathbf{u},
\end{equation}
which is subject to
\begin{gather}
    \mathbf{u}(y=\pm1)=\mathbf{0},\\
    \nabla\cdot\mathbf{u}=0,
\end{gather}
with $\mathbf{M}(\mathbf{u})=-(\mathbf{u}_b\cdot\nabla)\mathbf{u}-(\mathbf{u}\cdot\nabla)\mathbf{u}_b-(\mathbf{u}\cdot\nabla)\mathbf{u}$. The influence matrix (IM) algorithm has been developed to numerically advance this particular type of dynamical systems, which have a Laplacian linear term and gradient of a scalar on the right-hand side, and are subject to zero-divergence constraint and homogeneous Dirichlet BCs at the walls. This algorithm enforces zero divergence and the homogeneous Dirichlet BCs within the time-stepping process while the scalar field is handled implicitly and is not resolved as a separate variable (\cite{Kleiser1980}; \cite{Canuto2007}, \textsection{3.4}).
We use the IM algorithm, and introduce the following five steps which advance $\mathbf{u}$ under the variational dynamics \eqref{eq:adjoint_evolution_divergence_free}-\eqref{eq:adj_P_BC_} for one time step of size $\Delta\tau$:

\begin{enumerate}
    \item The current velocity field $\mathbf{u}$, that satisfies $\nabla\cdot\mathbf{u}=0$ and $\mathbf{u}(y=\pm1)=\mathbf{0}$, is advanced under the NSE dynamics for one physical time step $\Delta t$ using the IM algorithm. This yields the updated velocity $\mathbf{u}^{\Delta t}$ where the IM algorithm ensures $\nabla\cdot\mathbf{u}^{\Delta t}=0$ and $\mathbf{u}^{\Delta t}(y=\pm1)=\mathbf{0}$.
    
    \item The residual field $\mathbf{r}_1$, which is by definition the right-hand side of the NSE \eqref{eq:momentum}, is approximated via finite differences
    \begin{equation}
    \label{eq:r1_approximation}
        \mathbf{r}_1=\dfrac{\partial\mathbf{u}}{\partial t}\approx\dfrac{\mathbf{u}^{\Delta t}-\mathbf{u}}{\Delta t}.
    \end{equation}
    Since both $\mathbf{u}$ and $\mathbf{u}^{\Delta t}$ are divergence-free and satisfy homogeneous Dirichlet BCs at the walls, $\nabla\cdot\mathbf{r}_1=0$ and $\mathbf{r}_1(y=\pm1)=\mathbf{0}$.

    \item The current residual field $\mathbf{r}_1$ is advanced under the dummy dynamics \eqref{eq:g_trick}-\eqref{eq:g_trick_zero_divergence} for one time step $\Delta\hat{\tau}$ using the IM algorithm, which yields $\mathbf{r}_1^{\Delta\hat{\tau}}$. The IM algorithm ensures that $\nabla\cdot\mathbf{r}_1^{\Delta\hat{\tau}}=0$ and $\mathbf{r}_1^{\Delta\hat{\tau}}(y=\pm1)=\mathbf{0}$.
    
    \item The right-hand side of Equation \eqref{eq:g_trick} is approximated via finite differences
    \begin{equation}
    \label{eq:f_approximation}
        \mathbf{f}=\dfrac{\partial\mathbf{r}_1}{\partial\hat{\tau}} \approx \dfrac{\mathbf{r}_1^{\Delta \hat{\tau}}-\mathbf{r}_1}{\Delta \hat{\tau}}.
    \end{equation}
    Since both $\mathbf{r}_1$ and $\mathbf{r}_1^{\Delta\hat{\tau}}$ are divergence-free and satisfy homogeneous Dirichlet BCs at the walls, $\nabla\cdot\mathbf{f}=0$ and $\mathbf{f}(y=\pm1)=\mathbf{0}$.

    \item Having approximated $\mathbf{f}$, which is the descent direction at the current fictitious time $\tau$, we advance the velocity for one step of size $\Delta\tau$ using
    \begin{equation}
        \label{eq:forward_Euler_update}
        \mathbf{u}^{\Delta\tau} = \mathbf{u} + \Delta\tau\,\mathbf{f}.
    \end{equation}
    Since both $\mathbf{u}$ and $\mathbf{f}$ are divergence-free and take zero value at the walls, the updated velocity satisfies $\nabla\cdot\mathbf{u}^{\Delta\tau}=0$ and $\mathbf{u}^{\Delta\tau}(y=\pm1)=\mathbf{0}$.
\end{enumerate}

The finite differences \eqref{eq:r1_approximation} and \eqref{eq:f_approximation} affect the accuracy of time-stepping the variational dynamics, but they do not interfere with imposing the boundary condition $\mathbf{u}(y=\pm1)=\mathbf{0}$ and the constraint $\nabla\cdot\mathbf{u}=0$ within machine precision.
The low accuracy of the first-order finite differences does not affect the accuracy of the obtained equilibrium solution since both $\mathbf{r}_1$ and $\mathbf{f}$ tend to zero when an equilibrium is approached.
We are also not concerned about the low accuracy of the first-order forward Euler update rule \eqref{eq:forward_Euler_update} since the objective is to obtain the attracting equilibria of the adjoint-descent dynamics reached at $\tau\to\infty$.
Therefore, the introduced procedure is able to construct equilibrium solutions within machine precision.

We implement this procedure in \textit{Channelflow 2.0}, an open-source software package for numerical analysis of the incompressible NSE in wall-bounded domains. 
In this software, an instantaneous divergence-free velocity field is represented by Chebyshev expansion in the wall-normal direction $y$ and Fourier expansion in the periodic directions $x$ and $z$:
\begin{equation}
\label{eq:Foureir_Chebyshev_Fourier}
    u_j(x,y,z)=\sum_{\substack{m,p\in\mathbb{Z}\\n\in\mathbb{W}}}\hat{u}_{m,n,p,j}T_n(y)e^{2\pi i\left({mx}/{L_x}+{pz}/{L_z}\right)}\;;\quad j=1,2,3,
\end{equation}
where $T_n(y)$ is the $n$-th Chebyshev polynomial of the first kind, $i$ is the imaginary unit, and indices 1 to 3 specify directions $x$, $y$ and $z$, respectively.
\textit{Channelflow 2.0} employs the influence matrix algorithm for time-marching the NSE \eqref{eq:u_trick}.
With modification for the nonlinear term $\mathbf{N}(\mathbf{r}_1)$, Equation \eqref{eq:g_trick} can also be advanced in time.

\section{Application to plane Couette flow}
\label{sec:results}
We apply the introduced variational method to plane Couette flow (PCF), the flow between two parallel plates moving at equal and opposite velocities. PCF is governed by the general NSE \eqref{eq:momentum}-\eqref{eq:NSE_BC3} with the laminar base flow $\mathbf{u}_b=[y,0,0]^\top$. Due to the periodicity in $x$ and $z$, PCF is equivariant under continuous translations in these directions:
\begin{equation}
    \tau(\ell_x, \ell_z):\;\left[u,v,w\right](x,y,z) \mapsto \left[u,v,w\right](x+\ell_x,y,z+\ell_z),
\end{equation}
where $u$, $v$ and $w$ are the components of $\mathbf{u}$ in $x$, $y$ and $z$ directions, respectively.
In addition, PCF is equivariant under two discrete symmetries as well, rotation around the line $x=y=0$:
\begin{equation}
    \sigma_1:\;\left[u,v,w\right](x,y,z) \mapsto \left[-u,-v,w\right](-x,-y,z),
\end{equation}
and reflection with respect to the plane $z=0$:
\begin{equation}
    \sigma_2:\;\left[u,v,w\right](x,y,z) \mapsto \left[u,v,-w\right](x,y,-z).
\end{equation}
The variational dynamics \eqref{eq:adjoint_evolution_divergence_free}-\eqref{eq:adj_P_BC_} is easily verified to be equivariant under the same continuous and discrete symmetry operators. Therefore, the variational dynamics preserves these symmetries, if present in the initial condition. In the following, we demonstrate the convergence of multiple equilibrium solutions from guesses both within a symmetry-invariant subspace and outside.

\subsection{Results}
We search for equilibria of PCF at $Re=400$ within a domain of dimensions $L_x=2\pi/1.14$ and $L_z=2\pi/2.5$ (see \textsection{\ref{sec:governing_equations}}). This domain was first studied by \cite{Waleffe2002}. Several equilibrium solutions of PCF in this domain at $Re=400$ were computed by \cite{Gibson2008,Gibson2009}. These are available in the database on \href{http://channelflow.org}{channelflow.org}. Here, the flow field is discretised with $N_y=31$ collocation points in the wall-normal direction and $N_x=N_z=32$ points in the lateral directions.
The adjoint-descent dynamics is numerically integrated by the forward Euler scheme \eqref{eq:forward_Euler_update} with $\Delta\tau=0.03$, and $\mathbf{r}_1$ and $\mathbf{f}$ are approximated via finite differences \eqref{eq:r1_approximation} and \eqref{eq:f_approximation} with the step size $\Delta t=0.25$ and $\Delta\hat{\tau}=0.25$, respectively (see \textsection{\ref{sec:numerical_implementation}}). An accurate finite-difference approximation of $\mathbf{r}_1$ and $\mathbf{f}$ suggests choosing $\Delta t$ and $\Delta\hat{\tau}$ as small as possible. However, smaller values for these step sizes result in a less stable forward Euler integration scheme, requiring a smaller value of $\Delta\tau$ to remain stable. 
Since for an equilibrium solution $\mathbf{r}_1=\mathbf{f}=\mathbf{0}$, larger values of $\Delta t$ and $\Delta\hat{\tau}$ do not diminish the accuracy of the obtained equilibrium solution. Consequently, we empirically choose values for $\Delta t$ and $\Delta\hat{\tau}$ so that a reasonably large value for $\Delta \tau$ can be used.

To verify the scheme and its implementation, we converge the so-called `Nagata's lower branch' equilibrium solution (\cite{Nagata1990}) at $Re=400$.
As initial guess, we take an equilibrium solution on the same branch but at a significantly different $Re$.
The Nagata's lower branch solution at $Re=400$ continued from Nagata's original domain dimensions to the ones considered here is available in the database on \href{http://channelflow.org}{channelflow.org}.
We continue this equilibrium solution to $Re=230$, and use the resulting solution to initialise both the adjoint-descent variational method and the standard Newton iterations at $Re=400$.
The standard Newton iterations, i.e. without optimisations such as hook steps, fail to converge. However, the adjoint-descent variational method successfully converges to the equilibrium solution at $Re=400$ on the same branch.

\begin{figure}
     \centering
     \includegraphics[height = 5.5cm]{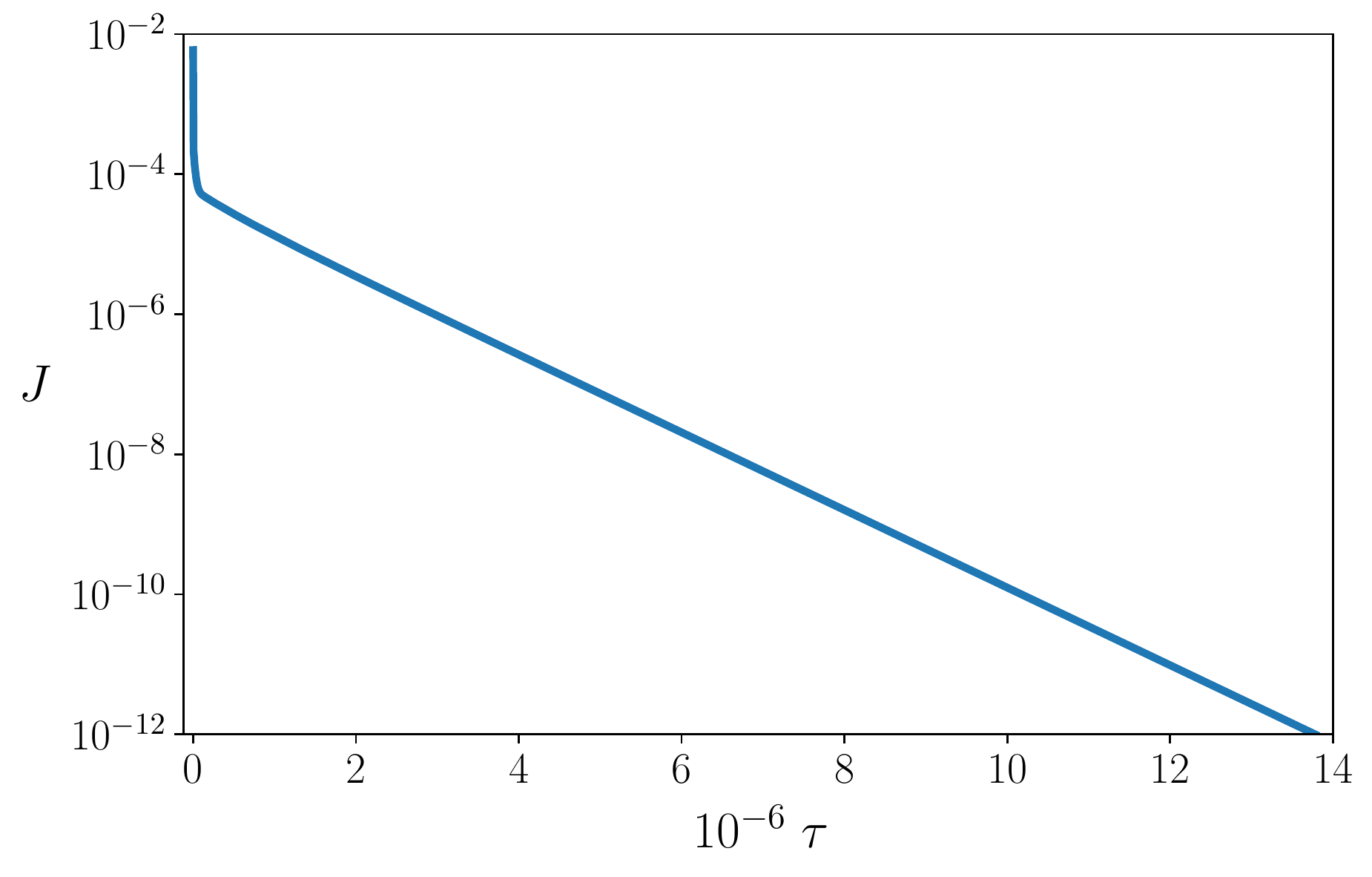}
     \caption{Convergence of the adjoint-descent variational method for constructing an equilibrium solution of the plane Couette flow. The minimisation of the cost function $J$ evolves the initial guess towards a true equilibrium solution at which $J=0$.}
\label{fig:norm}
\end{figure}

\begin{figure}
     \centering
     \includegraphics[height = 6.5cm]{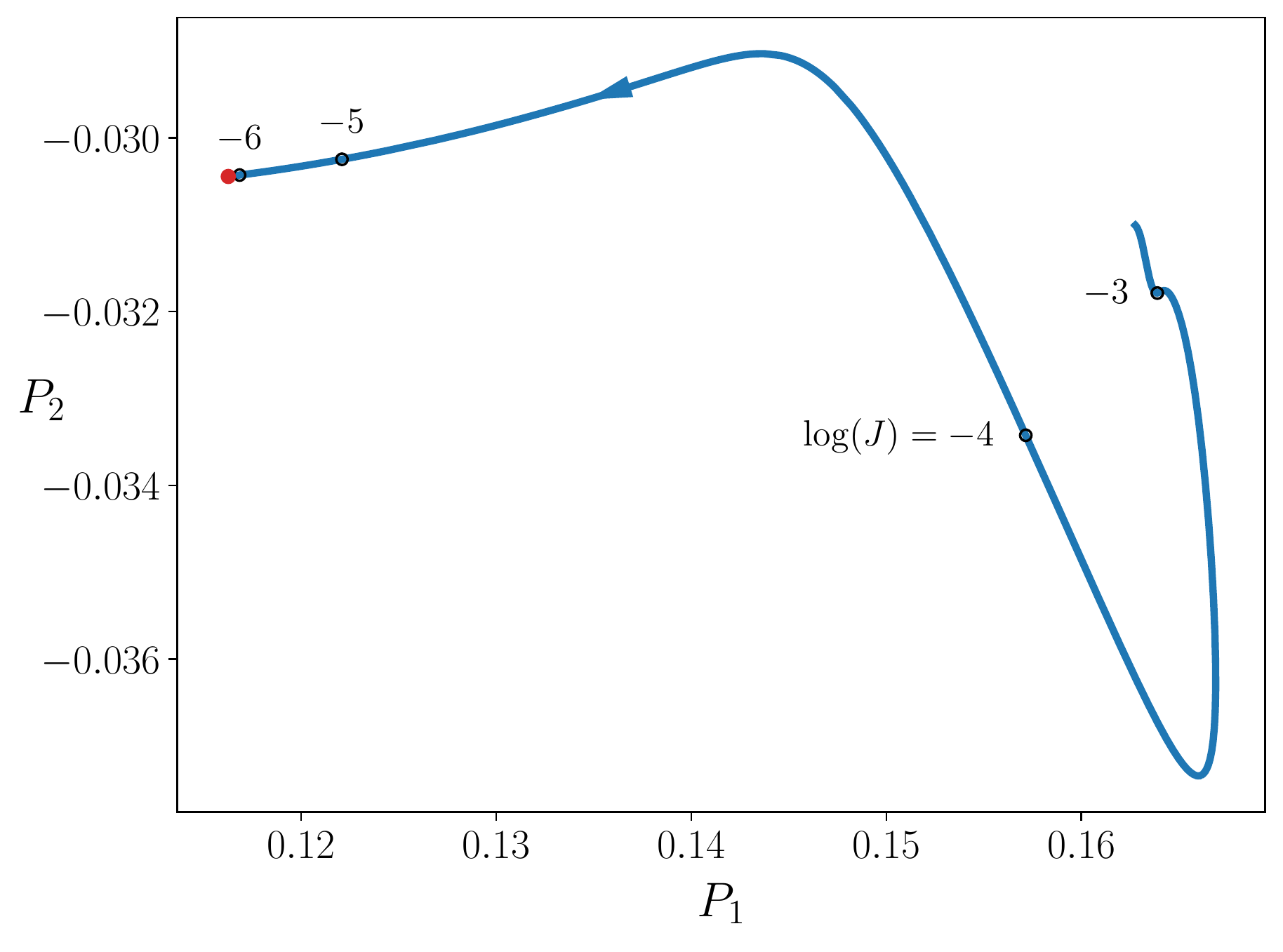}
     \caption{The trajectory of the adjoint-descent dynamics along which the cost function $J$ decreases monotonically as shown in Figure \ref{fig:norm}. The projection shows $P_2=\Re\{\hat{u}_{0,5,0,1}\}$ against $P_1=\Re\{\hat{u}_{0,3,0,1}\}$. The majority of the trajectory is traversed rapidly at the beginning, as indicated by a sharp drop of $J$ in Figure \ref{fig:norm}, followed by a slow traversal of the remaining portion towards the asymptotic solution, reflected in Figure \ref{fig:norm} as an exponential decay of the cost function.}
\label{fig:state_space}
\end{figure}

Along the trajectory of the adjoint-descent dynamics, the cost function initially drops rapidly and subsequently decreases with an exponential rate, as shown in Figure \ref{fig:norm}. The exponential decrease of the cost function is explained by the dynamical system picture of the adjoint descent: the adjoint-descent dynamics converges to a stable fixed point, hence the evolution is dominated by the slowest eigenmode of the linearised dynamics in the vicinity of that fixed point.
The sharp initial drop and the following exponential decay of the cost function are reflected in fast and slow traversal, respectively, of the trajectory within the state space. Figure \ref{fig:state_space} presents a 2D projection of the trajectory, with markers indicating that the majority of the trajectory is traversed quickly in the beginning of the integration, and the majority of the integration time is spent on the remaining, much shorter portion of the trajectory. For instance, the portion of the trajectory traversed during the first $1.2\times10^6$ fictitious time units, that decreases the cost function from $J=5.9\times10^{-3}$ to $J=10^{-5}$, is considerably longer than the remaining portion which takes over $90\,\%$ of the integration time to be traversed. In Figure \ref{fig:state_space}, $P_1$ and $P_2$ are the real parts of $\hat{u}_{0,3,0,1}$ and $\hat{u}_{0,5,0,1}$, i.e. the coefficients of the third and the fifth Chebyshev polynomial in the expansion of the mean streamwise velocity in $y$ (see Equation \eqref{eq:Foureir_Chebyshev_Fourier}). The visualisation of the trajectory in different projections of the state space yields a similar observation.

Nagata's lower branch equilibrium solutions are symmetric under shift-and-rotate symmetry $s_1=\tau(L_x/2,L_z/2)\sigma_1$:
\begin{equation}
    s_1\left[u,v,w\right](x,y,z) = \left[-u,-v,w\right](-x+L_x/2,-y,z+L_z/2),
\end{equation}
and shift-and-reflect symmetry $s_2=\tau(L_x/2,0)\sigma_2$:
\begin{equation}
    s_2\left[u,v,w\right](x,y,z) = \left[u,v,-w\right](x+L_x/2,y,-z).
\end{equation}
Therefore, the initial guess in the present example, namely the Nagata's lower branch solution at $Re=230$, is symmetric under $s_1$ and $s_2$ that are preserved by the adjoint-descent dynamics.
The velocity field remains symmetric under $s_1$ and $s_2$ without explicitly enforcing them during the forward integration until the equilibrium solution on the same branch at $Re=400$ is converged.

\begin{figure}
     \centering
     \includegraphics[height = 6.5cm]{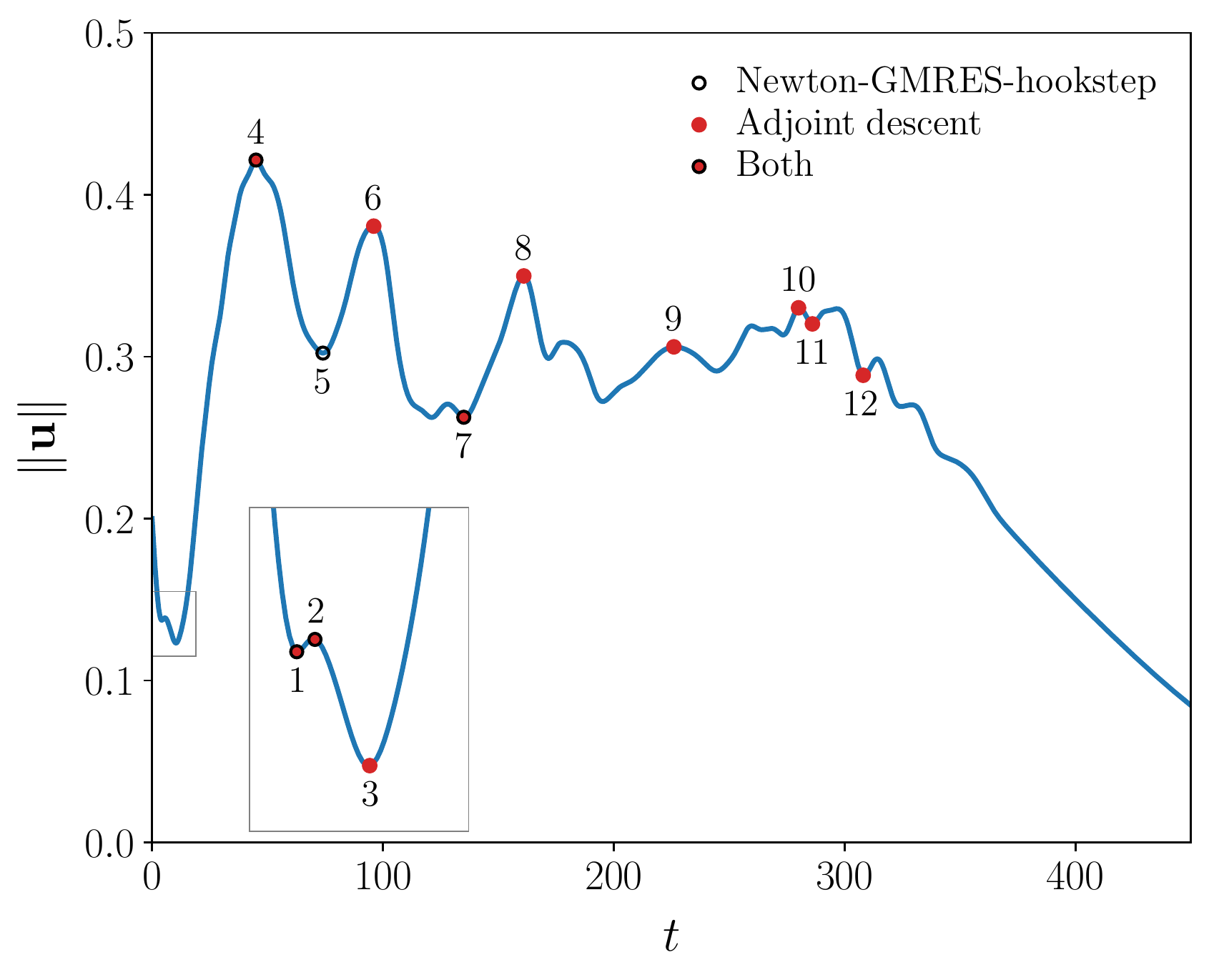}
     \caption{The $L_2$-norm of the velocity field against the physical time $t$ in direct numerical simulation from a random initial condition. The snapshots corresponding to the local extrema of $\|\mathbf{u}\|$ are selected as guesses for an equilibrium solution. Table \ref{table:equilibria} summarises the result of the convergence from each guess using Newton-GMRES-hookstep and the adjoint-descent variational method.}
\label{fig:DNS}
\end{figure}

To further investigate the robustness of the adjoint-descent variational method in successfully converging from inaccurate guesses, we initialise the method with guesses obtained from a direct numerical simulation.
We construct a random divergence-free velocity field with $L_2$-norm $\|\mathbf{u}\|=0.2$, and time-march the NSE along a turbulent trajectory until the flow laminarises. 
The initial condition and therefore the entire trajectory are not symmetric under any of the symmetries allowed by the PCF.
We extract the local extrema of $\|\mathbf{u}\|$ as a function of time $t$, where $\partial\|\mathbf{u}\|/\partial t=0$, as guesses for potential equilibrium solutions.
Figure \ref{fig:DNS} shows $\|\mathbf{u}\|$ plotted against $t$ from which $26$ guesses are extracted.
The standard Newton iterations do not converge starting from any of the guesses. With hook-step optimisation, $5$ of the searches converge within $50$ Newton-GMRES-hookstep (NGh) iterations.
The converged solutions include the trivial laminar solution $\mathbf{u}=\mathbf{0}$ as well as two nontrivial solutions EQ1 and EQ3 (see Tables \ref{table:equilibria} and \ref{table:equilibria_properties} for properties of the converged solutions).
By integrating the adjoint-descent dynamics, $11$ of the guesses converge to an equilibrium solution. These solutions include the trivial solution as well as five nontrivial equilibria EQ1 to EQ5 (see Tables \ref{table:equilibria} and \ref{table:equilibria_properties}).
Among these solutions, EQ1, EQ4 and EQ5 have been documented in the literature (\cite{Gibson2009}). Yet, to the best of our knowledge, the equilibria labelled EQ2 and EQ3 have not been previously reported.
Snapshots that lead to a successful search via either NGh iterations or the adjoint-descent algorithm are marked on Figure \ref{fig:DNS}.

\begin{table}
    \begin{center}
        \def~{\hphantom{0}}
        \begin{tabular}{c | cc | c}
            snapshot & NGh iterations & NGh solution & adjoint-descent solution\\[3pt]
            1 & 13 & EQ0 & EQ0 \\
            2 & 11 & EQ0 & EQ0 \\
            3 & - & - & EQ0 \\
            4 & 23 & EQ1 & EQ2 \\
            5 & 15 & EQ1 & - \\
            6 & - & - & EQ1 \\
            7 & 13 & EQ3 & EQ2 \\
            8 & - & - & EQ4 \\
            9 & - & - & EQ3 \\
            10 & - & - & EQ5 \\
            11 & - & - & EQ5 \\
            12 & - & - & EQ3 \\
        \end{tabular}
        \caption{The list of the equilibrium solutions converged by Newton-GMRES-hookstep (NGh) and the adjoint-descent variational method from the guesses marked in Figure \ref{fig:DNS}. See Table \ref{table:equilibria_properties} for properties of the equilibria EQ0 to EQ5.}
        \label{table:equilibria}
    \end{center}
\end{table}

\begin{table}
    \begin{center}
        \def~{\hphantom{0}}
        \begin{tabular}{ccc}
            solution & $\|\mathbf{u}\|$ & $D/D_\mathrm{lam}$\\[3pt]
            EQ0 & 0 & 1 \\
            EQ1 & 0.385858 & 3.04427 \\
            EQ2 & 0.268277 & 1.76302 \\
            EQ3 & 0.240519 & 1.60348 \\
            EQ4 & 0.168131 & 1.45374 \\
            EQ5 & 0.328654 & 2.37353 \\
        \end{tabular}
        \caption{Properties of the equilibrium solutions converged by Newton-GMRES-hookstep and the adjoint-descent variational method (see Table \ref{table:equilibria} and Figure \ref{fig:DNS}). The second column contains the $L_2$ norm of the solutions and the third column contains the total energy dissipation of the solutions normalised by that of the laminar base flow.}
        \label{table:equilibria_properties}
    \end{center}
\end{table}

The variational method succeeds in more than twice as many cases as the NGh method, and extracts three more non-trivial equilibria from a turbulent trajectory with a crude criterion for selecting guesses.
This suggests that the basin of attraction to converge an equilibrium solution is typically larger for the adjoint-descent variational method compared to the NGh method.
However, the larger basin of attraction does not necessarily contain the smaller one. Notice, for instance, that the NGh iterations and the adjoint-descent algorithm converge to different equilibrium solutions when initialised with the snapshot 4, or the NGh iterations converge when initialised with the snapshot 5 while the adjoint-descent does not.

Despite the advantage of the variational method in successfully converging from inaccurate guesses, this method exhibits a very slow rate of convergence.
For instance, the convergence in our first example (Figures \ref{fig:norm}) takes near $650$ hours of wall clock time on one core of a 2.60GHz Intel Xeon E5-2640 CPU.
Besides the improvements on the computer programming side, such as parallel computations on multiple CPU cores, the convergence can be significantly accelerated by employing the inherent predictability of the variational dynamics, namely its almost linear behaviour when the trajectory reaches the vicinity of a solution.
In the following, we introduce a data-driven technique for such an acceleration.

\section{Accelerating the convergence}
\label{sec:accelerating}
The variational dynamics evolves along the gradient descent of the cost function. As a result, this dynamics is globally contracting, and almost all its trajectories eventually converge to a stable fixed point where the cost function takes a minimum value.
When the trajectory of the adjoint-descent dynamics has got sufficiently close to its destination fixed point, the cost function is well represented by a quadratic function and its gradient flow is almost linear. The approximately linear behaviour of the variational dynamics in the vicinity of an asymptotic fixed point inspires the idea of the following data-driven technique for accelerating the slow convergence of the variational method.

Our acceleration technique aims to approximate the expected linear dynamics and thereby approximate the equilibrium solution of the adjoint-descent dynamics. Since the destination fixed point is not known a priori, linearisation around the unknown fixed point is obviously not possible. Instead, we employ dynamic mode decomposition (DMD) to approximate the linear dynamics based on the available portion of the trajectory that has been traversed.
DMD is a regression framework that constructs the best-fit linear model over a series of snapshots (\cite{Rowley2009,Schmid2010,Kutz2016,Schmid2022}). The equilibrium solution of the adjoint-descent dynamics is approximated by letting the fictitious time go to infinity in the approximated linear system.

\subsection{Dynamic mode decomposition (DMD)}
Suppose each instantaneous spatially resolved flow field $\mathbf{u}(\mathbf{x};\tau)$ is represented by an $N$-dimensional real-valued column vector $\psi(\tau)$. $M$ snapshots $\psi_k=\psi(\tau_k);\;k=1,\dots,M$ along a single trajectory can be related to the snapshots taken $\delta\tau$ later along the same trajectory, $\psi'_k=\psi(\tau_k+\delta \tau)$, via the following linear relation:
\begin{equation}
    \psi'_k = \mathbf{A}\psi_k + e_k;\quad k=1,\dots,M,
\end{equation}
where $e_k$ is the error in approximating $\psi'_k$ by the linear map $\psi_k\mapsto\mathbf{A}\psi_k$. DMD constructs the ${N\times N}$ linear operator $\mathbf{A}$ which minimises the sum of squares of the elements of $e_k$ over all $M$ snapshot pairs:
\begin{equation}
\label{eq:def_DMD_matrix}
    \mathbf{A} := \Psi'\Psi^+,
\end{equation}
where $\Psi:=\big[\psi_1 \;\; \psi_2 \;\; \dots \;\; \psi_M\big]$, $\Psi':=\big[\psi'_1 \;\; \psi'_2 \;\; \dots \;\; \psi'_M\big]$, and the superscript $+$ denotes the Moore–Penrose pseudo-inverse. The dimensionality of the system can be prohibitively large for constructing $\mathbf{A}$ directly as defined in Equation \eqref{eq:def_DMD_matrix}, which is typically the case in a fluid dynamics problem. Therefore, we instead use a rank-reduced representation of this matrix. For this, the data matrix $\Psi$ is factorised via singular value decomposition (SVD) as $\Psi\approx\mathbf{U}\Sigma\mathbf{V}^\top$ with truncation rank $r$. The $r\times r$ projection of $\mathbf{A}$ on the POD modes $\mathbf{U}$ is
\begin{equation}
    \tilde{\mathbf{A}}=\mathbf{U}^\top\mathbf{A}\mathbf{U}=\mathbf{U}^\top\Psi'\mathbf{V}\Sigma^{-1}.
\end{equation}
The dynamic modes and their temporal behaviour are constructed from the eigendecomposition of $\tilde{\mathbf{A}}$: Dynamic modes are $\phi_q=\left(\Psi'\mathbf{V}\Sigma^{-1}\right)v_q$ with $q=1,\dots,r$, where $v_q$ are eigenvectors of $\tilde{\mathbf{A}}$; and the dynamic mode $\phi_q$ evolves as $e^{\omega_q\tau}$ where $\omega_q=\ln(\lambda_q)/\delta\tau$ and $\lambda_q$ is the eigenvalue of $\tilde{\mathbf{A}}$ associated with $v_q$.
Finally, the linear evolution of $\psi(\tau)$ is approximated as
\begin{equation}
\label{eq:DMD_linear_approximation}
    \psi(\tau) \approx \sum_{q=1}^{r}b_q\phi_q e^{\omega_q \tau},
\end{equation}
where $b_q$ are the amplitudes of the dynamic modes at a reference time, for instance at $\tau_M$.
Based on this linear model we approximate the asymptotic equilibrium solution of the variational dynamics as follows.

\subsection{Numerical implementation}
In order to approximate the linear dynamics using DMD, we collect snapshots $\psi_k$ after the initial fast drop in the cost function, when it decays exponentially.
The exponential decay implies that the dynamics is almost linear and dominated by only few of the slowest attracting eigenmodes.
As a result, the snapshot matrices are of low column rank.
In the vicinity of the yet to-be-found attracting fixed point, the Jacobian of the descent dynamics is the negative of the (discretised) second variation, or Hessian, of the cost function, and thus symmetric. Consequently, the eigenvalues of the linear dynamics approximated by the DMD are real.
Suppose the dynamic modes are sorted in increasing order of $|\omega_q|$. Then, the exponent $\omega_1$ is significantly closer to zero than the rest, and $\omega_2,\dots,\omega_r$ are negative.
By assuming $\omega_1\approx 0$, the linear model \eqref{eq:DMD_linear_approximation} can be expressed as the superposition of the steady state $\psi_s:=b_1\phi_1$, and the decaying terms $b_q\phi_q\exp(\omega_q\tau);\,q=2,\dots,r$. The steady state $\psi_s$ approximates the equilibrium solution of the almost linear adjoint-descent dynamics.
The state vector $\psi_s$ is mapped back to the corresponding flow field, from where the integration of the adjoint-descent dynamics is restarted.
Let $r^\ast$ denote the rank of the snapshot matrices. Then, the truncation rank $r\leq r^\ast$ is chosen such that the cost function associated with the approximated equilibrium is the smallest. We consistently found the minimum for $r=r^\ast$.
In the following, we demonstrate the acceleration of the first test case presented in \textsection{\ref{sec:results}}.

\begin{figure}
     \centering
     \includegraphics[height = 5.5cm]{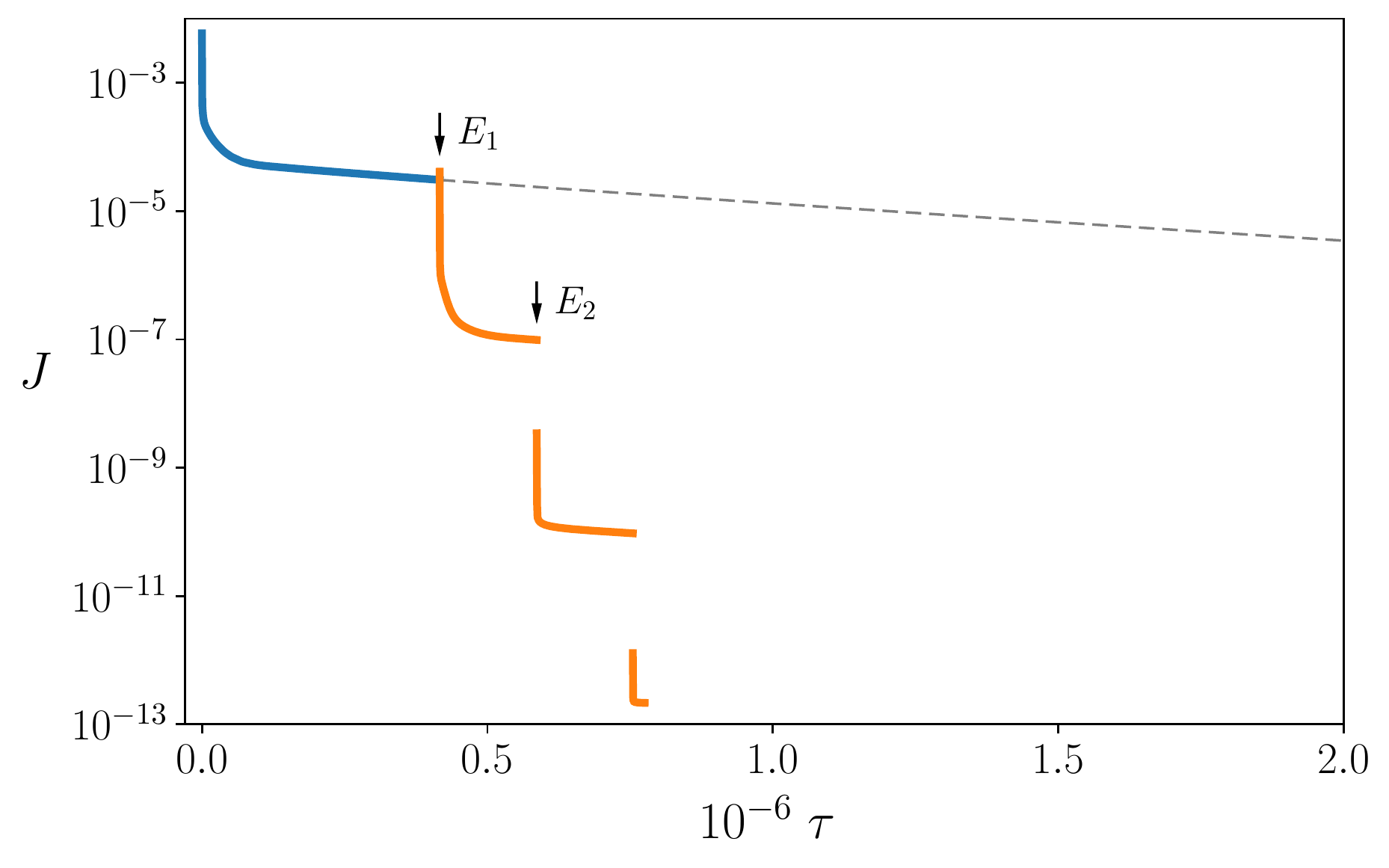}
     \caption{Acceleration of the convergence of the adjoint-descent variational method by successive DMD-based extrapolations. The extrapolation employs DMD to construct a best-fit linear model for the dynamics in the vicinity of an equilibrium, and approximates the asymptotic solution of the adjoint-descent dynamics by the asymptotic solution of the linear model.
     The acceleration technique reduces the total duration of the forward integration by $95\%$ in this example. The jumps in the state space associated with the first two extrapolations, $E_1$ and $E_2$, are shown in Figure \ref{fig:state_space_extrapolation}.}
\label{fig:norm_extrapolate}
\end{figure}

\begin{figure}
     \centering
     \includegraphics[height = 6.5cm]{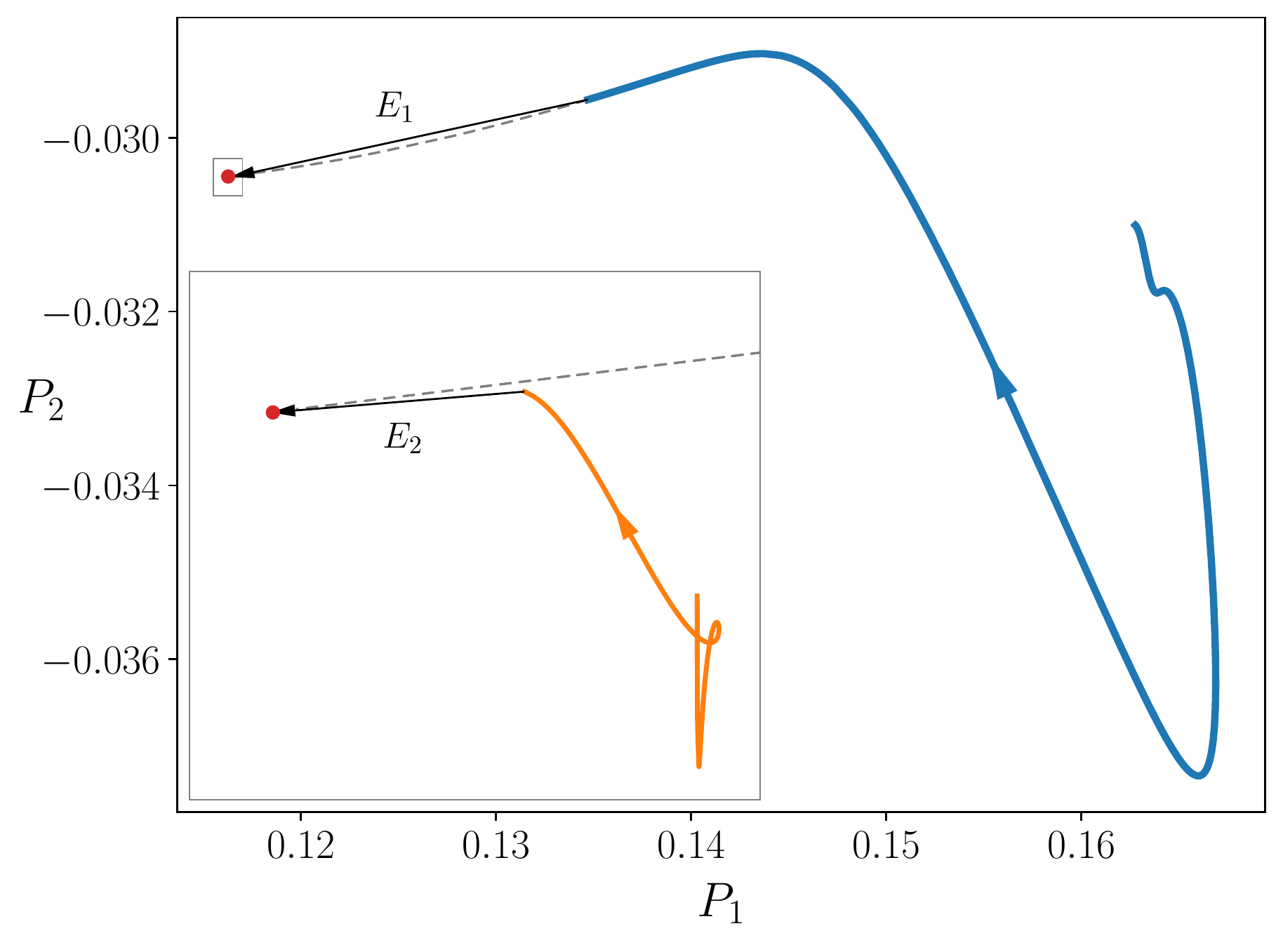}
     \caption{The trajectory of the accelerated adjoint-descent dynamics in the same 2D projection of Figure \ref{fig:state_space}. DMD-based extrapolations allow jumping to a state closer to the destination fixed point while avoiding integration of the adjoint-descent dynamics. The inset displays 225 times magnification of the area around the asymptotic solution.}
\label{fig:state_space_extrapolation}
\end{figure}

The snapshot vectors $\psi$ are the (real-valued) state vectors containing the minimum number of independent variables required for describing a divergence-free velocity field in Fourier-Chebyshev-Fourier spectral representation \eqref{eq:Foureir_Chebyshev_Fourier}. The vector $\psi$ has $N=20\,218$ elements for the discretisation used in \textsection{\ref{sec:results}}. 
Initially, we integrate the adjoint descent dynamics and let the cost function drop to $\log(J)=-4.5$ before performing the first DMD extrapolation.
The linear model is constructed using $M=100$ snapshots uniformly spaced over an interval of $2\times 10^4$ time units ($\delta\tau=200$). The next DMD extrapolations are performed using the same number of snapshots $M$ and the same spacing $\delta\tau$ while the adjoint dynamics is integrated forward in time for $15\times10^4$ time units before starting to collect new snapshots. The acceleration technique allows to achieve the convergence criterion $J=10^{-12}$ through $\tau=7.36\times10^5$ time units of total forward integration while without acceleration it takes $\tau=1.38\times10^7$ time units, that is almost $19$ times longer (see Figure \ref{fig:norm_extrapolate}, compare with Figure \ref{fig:norm}).
The time required for performing the extrapolation is negligible compared to the time required for the forward integration of the adjoint-descent dynamics.
The first DMD extrapolation has resulted in a slight increase in the value of $J$. The 2D projection of the state space, displayed in Figure \ref{fig:state_space_extrapolation}, shows that the first extrapolated state is significantly closer to the destination fixed point, despite being located on a higher level of $J$. By restarting the integration from the extrapolated state, the trajectory gets quickly attracted to the dominating eigendirection of the linearised dynamics resulting in a rapid drop in $J$ (see Figures \ref{fig:norm_extrapolate} and \ref{fig:state_space_extrapolation}).

Exploiting the linear behaviour of the variational dynamics, the acceleration technique typically achieves an order of magnitude speed-up in converging equilibria of PCF. The linear behaviour in the vicinity of an equilibrium solution at sufficiently large $\tau$ is a generic characteristic of the adjoint-descent variational method. Therefore, the introduced DMD-based acceleration technique is system-independent, and provided the snapshot vectors of the variational dynamics can be applied directly to any other problem.

\section{Summary and concluding remarks}
\label{sec:conclusion}
The unstable invariant solutions embedded within the chaotic attractor of the Navier-Stokes equations underpin the dynamics of a turbulent flow.
Despite the significance of invariant solutions for a dynamical description of chaotic flows, the identification of these solutions remains a computational challenge, demanding robust algorithms.
In this work, we have presented a matrix-free, adjoint-based variational method for computing equilibrium solutions of wall-bounded shear flows.
We have applied the introduced method to plane Couette flow, and demonstrated the convergence of multiple equilibrium solutions.
The variational method outperforms the state-of-the-art Newton iterations in successfully converging from inaccurate initial guesses, that suggests a larger basin of attraction.

The present method employs the norm of the right-hand side of the evolution equation as a cost function to penalise the deviation of a flow field from the equilibrium state. Thereby, the problem of finding an equilibrium solution is recast as the minimisation of the cost function.
To solve the minimisation problem, we adopted the variational approach of \cite{Farazmand2016} where the gradient of the cost function is constructed analytically via adjoint calculations, and thereby a matrix-free gradient descent method is utilised.
The cost function decreases monotonically along trajectories of the gradient descent dynamics until a minimum value is obtained.
The global minima of the cost function, taking zero value, correspond to the equilibrium solutions of the flow.
If a local minimum is obtained, the search for an equilibrium solution has failed. However, a local minimum of the cost function corresponds to the locally slowest state with respect to the chosen norm. This provides a means of characterising the so-called `ghost' of a saddle-node bifurcation (\cite{Strogatz2018}), which may influence the emerging spatiotemporal structures in chaotic flows (see, for example, \cite{Reetz2020b}, \textsection{3.1}).

The present work describes two key contributions: First, we apply the adjoint-based variational method to 3D wall-bounded flows. Previously, the variational approach had only been successfully applied to a 2D Kolmogorov flow in a doubly periodic domain without walls (\cite{Farazmand2016,Parker2022}).
The primary challenge in extending the variational method for computing equilibria to wall-bounded flows lies in handling the nonlinear, nonlocal pressure in the presence of solid walls.
To overcome this challenge, we have formulated the variational dynamics in a way that an explicit computation of pressure is avoided, allowing for application to 3D wall-bounded flows.
We demonstrated the variational method for plane Couette flow.

The second contribution is addressing the slow convergence of the adjoint-based variational method, that poses a challenge in practically utilising this method for 3D Navier-Stokes equations. 
We propose a data-driven technique for accelerating the convergence by extrapolating the asymptotic fixed point of the variational dynamics based on the traversed portion of its trajectory.
Since any trajectory of the variational dynamics converges to a stable fixed point, the dynamics behaves almost linearly when the trajectory has got close enough to the asymptotic solution.
The extrapolation technique takes advantage of this predictability, and approximates the best-fit linear dynamics using dynamic mode decomposition (DMD).
The asymptotic solution of the approximated linear system approximates the asymptotic solution of the variational dynamics.
This results in an order-of-magnitude speed-up in the overall duration of the forward integration required to converge to a solution within machine accuracy. 
The proposed acceleration technique is based on the generic properties of gradient descent minimisation, and is therefore independent of the physical system of study.

The variational dynamics have been derived for the deviation of the velocity field from the laminar base flow.
Consequently, an identical formulation and implementation directly translates to other wall-bounded flows such as plane Poiseuille flow (PPF) and asymptotic suction boundary layer (ASBL) flow as only the respective base velocity profiles in the variational dynamics \eqref{eq:adjoint_evolution_divergence_free}-\eqref{eq:adj_P_BC_} needs to be adapted.
However, due to the mean advection in PPF and ASBL, steady solutions in these flows take the form of travelling waves or relative equilibria, that are equilibria in a moving frame of reference.
The present method can be extended to compute travelling waves by expressing the NSE in a moving frame of reference. The speed of the Galilean transformation is an additional unknown in the cost function whose minimisation yields relative equilibrium solutions (\cite{Farazmand2016}).
The handling of the pressure and boundary conditions remains unchanged.

In the derivation of the adjoint-descent dynamics we assume the base velocity profile to be known and constant, which is the case when a fixed mean pressure gradient is imposed on the flow. For the alternative integral constraint of fixed mass flux, the base velocity profile, or more precisely its amplitude, needs to be determined together with the velocity and pressure perturbations.
As for the generalisation for travelling waves, the additional unknown can be included in the variational formulation without modifying the handling of pressure and boundary conditions.

The advantages of the adjoint-based variational method have inspired its application in computing other invariant sets, such as periodic orbits (\cite{Azimi2022,Parker2022}) and connecting orbits (\cite{Ashtari2023a}).
These methods view the identification of a periodic or connecting orbit as a minimisation problem in the space of space-time fields with prescribed behaviour in the temporal direction. They then employ a similar adjoint-based technique to solve the minimisation problem.
The robust convergence of these extensions has so far only been demonstrated in 2D flows in a doubly periodic domain and for 1D model systems.
Like in computing equilibria, dealing with pressure is the key challenge in formulating the adjoint-based variational method for computing periodic or connecting orbits in 3D wall-bounded flows. In our ongoing research, the next step is to extend the introduced algorithm to the computation of more complex invariant solutions in wall-bounded flows via extensions of the adjoint-based variational method.

\backsection[Acknowledgements]{Authors would like to thank Sajjad Azimi, Jeremy P. Parker, Moritz Linkmann, and Matthias Engel for insightful discussions.}

\backsection[Funding]{This research has been supported by the European Research Council (ERC) under the European Union’s Horizon 2020 research and innovation programme (grant agreement no. 865677).}

\backsection[Declaration of interests]{The authors report no conflict of interest.}

\backsection[Author ORCIDs]{\\
    O. Ashtari \href{https://orcid.org/0000-0001-5684-4681}{https://orcid.org/0000-0001-5684-4681}\\
    T. M. Schneider \href{https://orcid.org/0000-0002-8617-8998}{https://orcid.org/0000-0002-8617-8998}}

\appendix
\section{Derivation of the adjoint operator}
\label{app:adjoint_derivation}
\subsection{Directional derivative of the residual}
Using indicial notation to specify the $x$, $y$ and $z$ components of vector quantities by the indices $i=1,2,3$, respectively, we write the residual of the momentum and continuity equations as
\begin{gather}
    \label{eq:NSE_residual_indicial}
    r_{1,i} = -u_{b,j}\dfrac{\partial \ug_i}{\partial x_j} -\ug_j\dfrac{\partial u_{b,i}}{\partial x_j} -\ug_j\dfrac{\partial \ug_i}{\partial x_j} -\dfrac{\partial \pg}{\partial x_i} +\dfrac{1}{Re}\dfrac{\partial^2 \ug_i}{\partial x_j \partial x_j},\\
    \label{eq:continuity_residual_indicial}
    r_2=\dfrac{\partial \ug_j}{\partial x_j},
\end{gather}
where repeated indices imply Einstein summation convention. The directional derivative of the residual components, $r_{1,i}$ and $r_2$, along $\mathbf{G}=[\mathbf{g_1},g_2]$ is found directly from the definition:
\begin{equation}
    \begin{split}
        \mathscr{L}_{1,i}(\mathbf{U};\mathbf{G})= \lim_{\epsilon\to0}\frac{r_{1,i}(\mathbf{U}+\epsilon\mathbf{G})-r_{1,i}(\mathbf{U})}{\epsilon}
        =\;& - u_{b,j}\dfrac{\partial g_{1,i}}{\partial x_j}- g_{1,j}\dfrac{\partial u_{b,i}}{\partial x_j}- g_{1,j}\dfrac{\partial \ug_i}{\partial x_j} \\
        &- \ug_j\dfrac{\partial g_{1,i}}{\partial x_j}-\dfrac{\partial g_2}{\partial x_i}+\dfrac{1}{Re}\dfrac{\partial^2 g_{1,i}}{\partial x_j \partial x_j},
    \end{split}
\end{equation}
\begin{equation}
        \mathscr{L}_2(\mathbf{U};\mathbf{G})= \lim_{\epsilon\to0}\frac{r_2(\mathbf{U}+\epsilon\mathbf{G})-r_2(\mathbf{U})}{\epsilon} = \frac{\partial g_{1,j}}{\partial x_j}.
\end{equation}

\subsection{The adjoint operator}
To derive the adjoint operator of the directional derivative of the residual, $\mathscr{L}(\mathbf{U};\mathbf{G})$, we expand the inner product of $\mathscr{L}(\mathbf{U};\mathbf{G})$ and the residual $\mathbf{R}$ as follows:
\begin{align*}
     \left<\mathscr{L}(\mathbf{U};\mathbf{G}),\mathbf{R}\right> & =\int_\Omega{\left(\mathscr{L}_1\cdot\mathbf{r}_1 +\mathscr{L}_2r_2\right)}\text{d}\mathbf{x} \\
     & = \int_\Omega \begin{aligned}[t] \bigg[\bigg(&- u_{b,j}\dfrac{\partial g_{1,i}}{\partial x_j}- g_{1,j}\dfrac{\partial u_{b,i}}{\partial x_j} - g_{1,j}\dfrac{\partial \ug_i}{\partial x_j} - \ug_j\dfrac{\partial g_{1,i}}{\partial x_j} \\
     &-\dfrac{\partial g_2}{\partial x_i} + \dfrac{1}{Re}\dfrac{\partial^2 g_{1,i}}{\partial x_j \partial x_j}\bigg) r_{1,i}+\left(\frac{\partial g_{1,j}}{\partial x_j}\right)r_2\bigg] \text{d}\mathbf{x}.
     \end{aligned}
\end{align*}
Integrating by parts we have

\begin{align*}
     \int_{x_{j,\text{min}}}^{x_{j,\text{max}}}{u_{b,j}\dfrac{\partial g_{1,i}}{\partial x_j}r_{1,i}}\text{d}x_j=u_{b,j}g_{1,i}r_{1,i}\Big|_{x_j=x_{j,\text{min}}}^{x_{j,\text{max}}}-\int_{x_{j,\text{min}}}^{x_{j,\text{max}}}{\dfrac{\partial (u_{b,j}r_{1,i})}{\partial x_j}g_{1,i}}\text{d}x_j,
\end{align*}

\begin{align*}
     \int_{x_{j,\text{min}}}^{x_{j,\text{max}}}{\ug_j\dfrac{\partial g_{1,i}}{\partial x_j}r_{1,i}}\text{d}x_j=\ug_jg_{1,i}r_{1,i}\Big|_{x_j=x_{j,\text{min}}}^{x_{j,\text{max}}}-\int_{x_{j,\text{min}}}^{x_{j,\text{max}}}{\dfrac{\partial (\ug_jr_{1,i})}{\partial x_j}g_{1,i}}\text{d}x_j,
\end{align*}

\begin{align*}
     \int_{x_{i,\text{min}}}^{x_{i,\text{max}}}{\frac{\partial g_2}{\partial x_i}r_{1,i}}\text{d}x_i=g_2 r_{1,i}\Big|_{x_i=x_{i,\text{min}}}^{x_{i,\text{max}}}-\int_{x_{i,\text{min}}}^{x_{i,\text{max}}}{\dfrac{\partial r_{1,i}}{\partial x_i}g_2}\text{d}x_j,
\end{align*}

\begin{align*}
     \int_{x_{j,\text{min}}}^{x_{j,\text{max}}}{\dfrac{\partial^2 g_{1,i}}{\partial x_j\partial x_j}r_{1,i}}\text{d}x_j=\left[\frac{\partial g_{1,i}}{\partial x_j}r_{1,i}- g_{1,i}\frac{\partial r_{1,i}}{\partial x_j}\right]_{x_j=x_{j,\text{min}}}^{x_{j,\text{max}}}+\int_{x_{j,\text{min}}}^{x_{j,\text{max}}}{\dfrac{\partial^2 r_{1,i}}{\partial x_j\partial x_j}g_{1,i}}\text{d}x_j,
\end{align*}

\begin{align*}
     \int_{x_{j,\text{min}}}^{x_{j,\text{max}}}{\frac{\partial g_{1,j}}{\partial x_j}r_2}\text{d}x_j=g_{1,j} r_2\Big|_{x_j=x_{j,\text{min}}}^{x_{j,\text{max}}}-\int_{x_{j,\text{min}}}^{x_{j,\text{max}}}{\dfrac{\partial r_2}{\partial x_j}g_{1,j}}\text{d}x_j.
\end{align*}
For $\mathbf{U},\mathbf{R},\mathbf{G}\in\mathscr{P}_0$, the following boundary terms cancel out either due to the periodicity of $\mathbf{U}$, $\mathbf{R}$ and $\mathbf{G}$ in $x$ and $z$, or due to $\mathbf{g}_1(y=\pm1)=\mathbf{0}$:
\begin{align*}
    u_{b,j}g_{1,i}r_{1,i}\Big|_{x_j=x_{j,\text{min}}}^{x_{j,\text{max}}}=0,
\end{align*}

\begin{align*}
     \ug_jg_{1,i}r_{1,i}\Big|_{x_j=x_{j,\text{min}}}^{x_{j,\text{max}}}=0,
\end{align*}

\begin{align*}
     g_{1,i}\frac{\partial r_{1,i}}{\partial x_j}\Big|_{x_j=x_{j,\text{min}}}^{x_{j,\text{max}}}=0,
\end{align*}

\begin{align*}
     g_{1,j} r_2\Big|_{x_j=x_{j,\text{min}}}^{x_{j,\text{max}}}=0.
\end{align*}
Similarly, the other two boundary terms cancel out either due to the periodicity of $\mathbf{R}$ and $\mathbf{G}$ in $x$ and $z$, or due to $\mathbf{r}_1(y=\pm1)=\mathbf{0}$:
\begin{align*}
     g_2 r_{1,i}\Big|_{x_i=x_{i,\text{min}}}^{x_{i,\text{max}}}=0,
\end{align*}

\begin{align*}
     \frac{\partial g_{1,i}}{\partial x_j}r_{1,i}\Big|_{x_j=x_{j,\text{min}}}^{x_{j,\text{max}}}=0.
\end{align*}
We now rewrite the inner product as
\begin{align*}
     \left<\mathscr{L}(\mathbf{U};\mathbf{G}),\mathbf{R}\right> = \int_\Omega \bigg( &\dfrac{\partial (u_{b,j}r_{1,i})}{\partial x_j}-r_{1,j}\dfrac{\partial u_{b,j}}{\partial x_i}-r_{1,j}\dfrac{\partial \ug_j}{\partial x_i}+\dfrac{\partial (\ug_jr_{1,i})}{\partial x_j}\\
     & +\frac{1}{Re}\dfrac{\partial^2 r_{1,i}}{\partial x_j\partial x_j}-\frac{\partial r_2}{\partial x_i}\bigg)g_{1,i}\text{d}\mathbf{x} +\int_\Omega{\left(\dfrac{\partial r_{1,i}}{\partial x_i}\right)g_2}\text{d}\mathbf{x},
\end{align*}
that can be written in the vector form as
\begin{align*}
     \left<\mathscr{L}(\mathbf{U};\mathbf{G}),\mathbf{R}\right> = & \int_\Omega\bigg(\left(\nabla\mathbf{r}_1\right)\left(\mathbf{u}_b+\mathbf{\ug}\right)-\left(\nabla(\mathbf{u}_b + \mathbf{\ug})\right)^\top\mathbf{r}_1+\dfrac{1}{Re}\Delta\mathbf{r}_1+r_2\mathbf{r}_1-\nabla r_2\bigg)\cdot\mathbf{g}_1\text{d}\mathbf{x}\\
     &+\int_\Omega{\left(\nabla\cdot\mathbf{r}_1\right)g_2}\text{d}\mathbf{x}.
\end{align*}
By definition
\[
    \left<\mathscr{L}(\mathbf{U};\mathbf{G}),\mathbf{R}\right> = \left<\mathbf{G},\mathscr{L}^\dagger(\mathbf{U};\mathbf{R})\right> = \int_\Omega{\left(\mathscr{L}^\dagger_1\cdot\mathbf{g}_1 +\mathscr{L}^\dagger_2g_2\right)}\text{d}\mathbf{x},
\]
therefore, the components of $\mathscr{L}^\dagger(\mathbf{U};\mathbf{R})$ are obtained as
\begin{align}
    \mathscr{L}^\dagger_1 &= \left(\nabla\mathbf{r}_1\right)\left(\mathbf{u}_b+\mathbf{\ug}\right)-\left(\nabla(\mathbf{u}_b + \mathbf{\ug})\right)^\top\mathbf{r}_1+\dfrac{1}{Re}\Delta\mathbf{r}_1+r_2\mathbf{r}_1-\nabla r_2, \\ 
    \mathscr{L}^\dagger_2 & = \nabla\cdot\mathbf{r}_1.
\end{align}

\bibliographystyle{jfm}

\end{document}